\documentclass[journal,comsoc]{IEEEtran}
\usepackage[T1]{fontenc}

\usepackage{tabularx}
\usepackage{xcolor}
\usepackage{multirow}
\usepackage{booktabs}
\usepackage{array}
\newcolumntype{L}[1]{>{\raggedright\let\newline\\\arraybackslash\hspace{0pt}}m{#1}}
\newcolumntype{C}[1]{>{\centering\let\newline\\\arraybackslash\hspace{0pt}}m{#1}}
\newcolumntype{R}[1]{>{\raggedleft\let\newline\\\arraybackslash\hspace{0pt}}m{#1}}

\usepackage[english]{babel}
\usepackage{amsthm}
\usepackage{amssymb}

\newtheorem{theorem}{Theorem}
\newtheorem{lemma}{Lemma}

\theoremstyle{plain}
\newtheorem{proposition}{Proposition}

\theoremstyle{plain}
\newtheorem{corollary}{Corollary}

\theoremstyle{remark}

\usepackage{cite}
\usepackage[pagewise]{lineno}

\ifCLASSINFOpdf
   \usepackage[pdftex]{graphicx}
\else
   or other class option (dvipsone, dvipdf, if not using dvips). graphicx
   \usepackage[dvips]{graphicx}
   \graphicspath{{../eps/}}
   \DeclareGraphicsExtensions{.eps}
\fi
\usepackage[normalem]{ulem}

\usepackage{amsmath}
\usepackage[cmintegrals]{newtxmath}
\usepackage{algorithmic}
\usepackage{algorithm}

\usepackage{verbatim}
\usepackage[table]{xcolor}
\usepackage{array}
\usepackage{makecell}

\usepackage[pagewise]{lineno}
\usepackage{upgreek}

\ifCLASSOPTIONcompsoc
  \usepackage[caption=false,font=normalsize,uabelfont=sf,textfont=sf]{subfig}
\else
  \usepackage[caption=false,font=footnotesize]{subfig}
\fi
\begin{document}
 
\setcounter{figure}{0}
\renewcommand{\figurename}{Fig.}
\renewcommand{\thefigure}{\arabic{figure}}
\title{Sensing-Secure ISAC: Ambiguity Function Engineering for Impairing Unauthorized Sensing}

\author{Kawon~Han,~\IEEEmembership{Member,~IEEE,} Kaitao~Meng,~\IEEEmembership{Member,~IEEE,} and Christos~Masouros,~\IEEEmembership{Fellow,~IEEE}

\thanks{Manuscript received xx, 2025. K. Han and C. Masouros declare a relevant patent application: United Kingdom Patent Application No. 2500617.2. The work of K. Han has been supported in part by UKRI under Grant EP/Y035933/1. The work of K. Meng has been supported in part by UKRI under Grant EP/Y02785X/1. The work of C. Masouros has been supported by the Smart Networks and Services Joint Undertaking (SNS JU) project 6G-MUSICAL under Grant Agreement No. 101139176 and the European Union Horizon Europe (HORIZON) Marie Skłodowska-Curie Actions Doctoral Network (MSCA-DN) ISLANDS under grant agreement No 101120544. \textit{(Corresponding author: Kaitao Meng)}

K. Han and C. Masouros are with the Department of Electronic and Electrical Engineering, University College London, London, UK (emails: {kawon.han,c.masouros}@ucl.ac.uk). 

K. Meng is with the the Department of Electrical and Electronic Engineering, University of Manchester, Manchester, UK (emails:kaitao.meng@manchester.ac.uk).}}

\maketitle

\begin{abstract}
The deployment of integrated sensing and communication (ISAC) in wireless networks brings along unprecedented vulnerabilities to authorized passive sensing, necessitating the development of secure sensing solutions. Unlike traditional wireless communication, where data security can be enhanced through data encryption, sensing security is more challenging to achieve. This is because sensing parameters are embedded within the target-reflected signal leaked to unauthorized passive radar sensing eavesdroppers (Eve), implying that they can silently extract sensory information without prior knowledge of the information data. To overcome this limitation, we propose a novel sensing-secure ISAC framework that ensures secure target detection and estimation for the legitimate system, while obfuscating unauthorized sensing without requiring any prior knowledge of Eve. Specifically, by introducing artificial imperfections into the ambiguity function (AF) of ISAC signals, we introduce artificial ghost targets into Eve’s range profile which increase its range estimation ambiguity. In contrast, the legitimate sensing receiver (Alice) can suppress these AF artifacts using mismatched filtering, albeit at the expense of signal-to-noise ratio (SNR) loss. Employing an OFDM signal, a structured subcarrier power allocation scheme is designed to shape the secure autocorrelation function (ACF), inserting periodic peaks to mislead Eve’s range estimation and degrade target detection performance. To quantify the sensing security level, we introduce peak sidelobe level (PSL) and integrated sidelobe level (ISL) as key performance metrics. Additionally, we analyze the three-way trade-offs between communication, legitimate sensing, and sensing security, highlighting the impact of the proposed sensing-secure ISAC signaling on system performance. Furthermore, we formulate a convex optimization problem to maximize ISAC performance while guaranteeing a certain sensing security level. Numerical results validate the effectiveness of the proposed sensing-secure ISAC signaling, demonstrating its ability to degrade Eve’s target estimation while preserving Alice’s performance. 
\end{abstract}


%
\IEEEpeerreviewmaketitle

\section{Introduction}
%
%
%
%
\IEEEPARstart{I}ntegrated Sensing and Communication (ISAC) has emerged as a promising paradigm for next-generation wireless networks, enabling dual-functional wireless communication and radar sensing within a unified system. By sharing spectrum and hardware resources, ISAC enhances spectral efficiency and reduces system costs, making it a key enabler for applications such as autonomous driving, smart cities, and industrial automation \cite{liu2022integrated, dong2022sensing, valiulahi2023net}. The ultimate goal of deploying ISAC in cellular networks is to provide coordinated sensing services at an unprecedented scale, offering opportunities to make future wireless networks more connected and sustainable \cite{meng2024cooperative, meng2025cooperative, strinati2025toward, ISAC3}.

However, integrating sensing into communication networks introduces new vulnerabilities, particularly in \textit{sensing security}, where third parties may exploit the sensing functionality to gain knowledge of targets and environments, potentially exposing private information. Such unauthorized access could result in unauthorized tracking or identification \cite{qu2024privacy}, potentially leading to misuse by an adversary. Specifically, ISAC systems deployed for human sensing may leak personal information to a sensing eavesdropper (Eve), such as vital signs \cite{shah2019rf}, speech \cite{han2021vocal}, and hand gestures \cite{wan2014gesture}, posing a significant risk of privacy disclosure.

In previous generations of wireless networks, such threats were rarely considered, as these networks primarily focused on providing data communication services \cite{andrews2014will}, with sensing functionality not being a point of concern. Consequently, physical layer security (PLS) techniques for wireless networks have primarily focused on protecting data security from eavesdropping \cite{wang2018survey}. Additionally, communication security can be ensured through data encryption at the upper layers \cite{shiu2011physical}. \textcolor{black}{Following this direction, PLS techniques in ISAC have also primarily focused on communication security, designing ISAC systems to balance the performance trade-offs between sensing, communication, and communication secrecy rate \cite{bazzi2023secure, boljevic2024sum}}. Moreover, the radar sensing functionality in ISAC is utilized to assist communication PLS by providing information about potential Eve \cite{su2023sensing, wang2024sensing}.
 
Unlike conventional communication security, securing the sensing functionality in ISAC is more challenging, as unauthorized passive eavesdroppers can exploit transmitted ISAC signals to infer target information without actively engaging in the system. Furthermore, there is no data link to encrypt for sensing security, as passive radar Eve can utilize ISAC signals of opportunity as both reference and surveillance signals. Thus, achieving sensing security inevitably requires physical layer techniques that deceive Eve through obfuscating the target sensing channel \cite{cigno2022integrating}.

\subsection{Related Works}
Sensing security in ISAC remains an underexplored research area, despite its critical importance in safeguarding privacy and preventing unauthorized surveillance. In response to these demands, a few studies on secure wireless sensing based on channel state information (CSI) have been presented recently. The works in \cite{yao2023interference, ruan2024leveraging, staat2022irshield, shenoy2022rf} utilize additional hardware to modify the physical environment, thereby imposing artificial channels on Eve. In \cite{yao2023interference}, rotating fans and antennas are employed to alter the physical wireless propagation channel. Meanwhile, the studies in \cite{ruan2024leveraging, staat2022irshield} introduce reconfigurable intelligent surfaces (RIS) to generate artificial signal reflections in Eve’s received signals. Similarly, RF-Protect in \cite{shenoy2022rf} creates ghost targets to Eve by deploying reflectors near targets. However, these approaches may be overly complex due to additional hardware requirements and may significantly impact legitimate sensing performance.

Instead of physical channel control, randomized transmission techniques have recently been considered in \cite{hernandez2023scheduled, cominelli2024physical} to protect WiFi-based sensing. The work in \cite{hernandez2023scheduled} proposed a randomly scheduled transmission scheme using multiple antennas, which is highly resilient to Eve while still enabling legitimate sensing. However, this method requires multiple independent streams to be transmitted from different antennas, which can significantly degrade both legitimate sensing and communication performance. Similarly, the randomized beamforming approach in \cite{cominelli2024physical} aims to confuse Eve in estimating the target channel. However, it has an inherent limitation, as it does not take legitimate sensing performance into account.

Alternatively, pilot modifications to obfuscate Eve's channel estimation have been proposed in \cite{ghiro2022implementation, abanto2020stay, zhu2018tu}. They apply secret scrambled functions to pilot symbols, allowing only the legitimate sensing receiver, which possesses the secret code, to extract the true CSI related to targets. Similar approaches have been extended to multi-antenna systems, including controlling the phase difference between transmitting antennas \cite{wang2024multi} and spatial-temporal source-defined channel encryption \cite{luo2024mimocrypt, hu2024wishield}. The objective of this CSI obfuscation is to prevent Eve from estimating target-related CSI variations, as Eve typically leverages long training sequences (LTS) to estimate CSI in WiFi-based sensing. However, it is important to note that these pilot modification techniques may be circumvented by an unauthorized passive radar eavesdropper, which performs radar sensing based on a reference signal received directly from the transmitter \cite{berger2010signal}. This implies that Eve can still estimate target channels using the modified pilots as a reference signal, without relying on prior known LTS, thereby bypassing the CSI obfuscation intentionally designed by the transmitter.

Unlike CSI-based sensing, a major focus in ISAC research is to facilitate radar sensing using communication data signals \cite{liu2024ofdm, liu2025uncovering}, which is highly vulnerable to attacks from passive sensing Eve. In this regard, secure-sensing ISAC designs have been explored in \cite{ren2024secure, jia2024illegal, zou2024securing}. These studies address sensing security by optimizing the detection probability in a cell-free ISAC system \cite{ren2024secure} or minimizing the Cramér-Rao bound (CRB) of legitimate sensing \cite{jia2024illegal}, while simultaneously keeping Eve's detection probability low. Additionally, by incorporating radar mutual information, ISAC signaling designs that maximize legitimate sensing performance have been proposed in \cite{zou2024securing}, subject to constraints on Eve's performance and communication quality-of-service (QoS). However, these initial works have a fundamental limitation: they require Eve's CSI at the legitimate ISAC transmitter. This assumption may not be entirely realistic, as a passive radar eavesdropper remains silent and does not reveal its location.

\subsection{Motivations and Contributions}
In summary of related works, securing the sensing functionality in ISAC from attacks by passive sensing Eve remains largely underexplored in three key aspects of a passive radar system: 1) There is no data link to encrypt for sensing security.  
2) Eve exploits the ISAC transmitter (TX) signal as both the reference and surveillance signals to detect and estimate targets. 3) It is challenging to utilize Eve's CSI for designing secure-sensing ISAC signaling. Motivated by these fundamental challenges, we aim to develop a comprehensive Eve-agnostic sensing-secure ISAC framework that not only addresses these issues but also provides a theoretical analysis of the three-way trade-offs among communication, legitimate sensing, and sensing security.
 
In this paper, we present a novel sensing-secure ISAC framework that leverages \textit{ambiguity function (AF) shaping} to enhance sensing security against sensing Eve. The main idea stems from the fact that radar sensing performance is fundamentally characterized by the AF of the transmitted signal, which provides a means to control the performance of an unknown Eve. Furthermore, the knowledge gap of the ISAC TX signal between the legitimate sensing receiver (Alice) and Eve introduces a degree of freedom (DoF) in Alice's receiver design, allowing it to employ mismatched filtering to mitigate the AF sidelobes. Consequently, judiciously shaping the AF enables the protection of sensing functionality in ISAC while inherently compromising sensing and communication (S\&C) performance, leading to the three-way trade-offs among communication, legitimate sensing, and sensing security. The main contributions of this paper are summarized as follows:

\begin{itemize}
    \item We propose an AF engineering framework for ISAC signals to achieve sensing security. By introducing artificial targets into Eve’s range profile, we mislead its sensing while allowing the legitimate sensing receiver to eliminate them using mismatched filtering, albeit at the cost of SNR loss. To achieve this, we design a structured subcarrier power allocation scheme in OFDM that shapes the autocorrelation function (ACF), introducing periodic peaks to degrade Eve’s range estimation and target detection performance.

    \item We establish key performance metrics for sensing security and legitimate sensing. Specifically, we quantify sensing security using peak sidelobe level (PSL) and integrated sidelobe level (ISL), while measuring legitimate sensing performance using SNR loss caused by reciprocal filtering (RF). Under these metrics, we theoretically derive the impact of AF shaping on each performance aspect and provide comprehensive insights into the performance gap between Alice and Eve.

    \item Building on the theoretical foundation of AF shaping, we investigate the three-way trade-offs among communication, legitimate sensing, and sensing security, revealing how AF shaping influences ISAC system performance. Additionally, we formulate and solve an optimization problem to achieve a flexible balance between S\&C performance while ensuring a guaranteed sensing security level, thereby enabling sensing-secure ISAC signaling.
\end{itemize}

$Notations$: Boldface variables with lower- and upper-case symbols represent vectors and matrices, respectively. $\textbf{A} \in \mathbb{C}^{N \times M}$ and $\textbf{B} \in \mathbb{R}^{N \times M}$ denotes a complex-valued ${N \times M}$ matrix $\textbf{A}$ and a real-valued ${N \times M}$ matrix $\textbf{B}$, respectively. Also, $\mathbf{0}_{N}$, $\mathbf{1}_{N}$ and $\mathbf{I}_N$ denote an $N \times 1$ column vector of zeros, an $N \times 1$ column vector of ones, and a $N \times N$ identity matrix, respectively. $({\cdot})^{T}$, $({\cdot})^{H}$, and $({\cdot})^{*}$ represent the transpose, Hermitian transpose, and conjugate operators, respectively. ${\text{diag}({\textbf{a}})}$ denotes a diagonal matrix with diagonal entries of a vector $\textbf{a}$. The operators $\odot$ and $\oslash$ represent the Hadamard (element-wise) product and the element-wise division, respectively. $\mathbb{E}{[\cdot]}$ is the statistical expectation operator.

\section{System Model}
\begin{figure}[t!]
    \centering
    \subfloat{\includegraphics[scale=0.65]{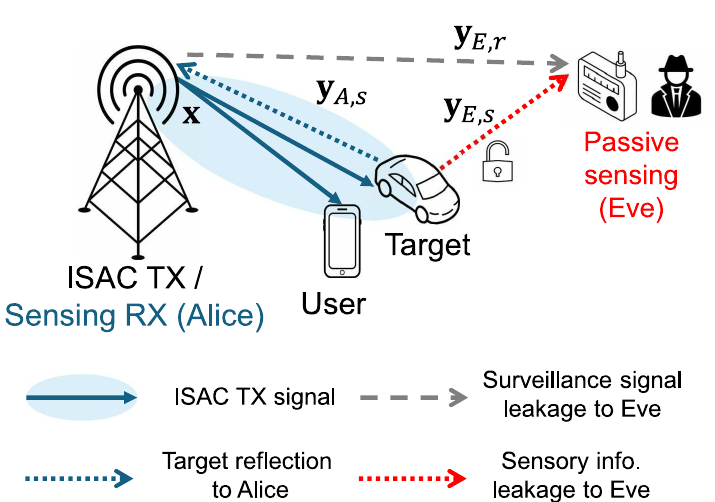}}
    \caption{Sensing-secure ISAC system to an unauthorized passive sensing eavesdropper. The leakage and target reflection to Eve cannot be controlled by the ISAC TX.}
    \label{f1}
\end{figure}
The proposed framework aims to develop a secure sensing ISAC system that prevents an unknown sensing eavesdropper (Eve) from detecting and estimating true targets by exploiting ISAC signals as signals of opportunity, as illustrated in Fig. \ref{f1}. We consider a single-antenna ISAC transmitter (TX) that transmits an orthogonal frequency division multiplexing (OFDM) signal. The legitimate sensing receiver (RX) (Alice) can either be collocated with the ISAC TX for monostatic sensing, as depicted in Fig. \ref{f1}, or be spatially separated from the TX for bistatic sensing. For both monostatic and bistatic sensing configuration, the legitimate sensing RX is assumed to have prior knowledge of the ISAC TX signal to demodulate the sensing channel.
Throughout this paper, we impose the following assumptions on the sensing Eve:
\begin{itemize}
    \item \textbf{(A.1)} The sensing Eve operates as a passive radar system equipped with a large number of receiving antennas or separately deployed directional antennas \cite{wojaczek2018reciprocal}, enabling perfect separation of a reference surveillance signal from the target echo signal.
    \item \textbf{(A.2)} The sensing Eve is aware of the location of the ISAC TX, including its range and angle with respect to Eve’s position. 
    \item \textbf{(A.3)} The sensing Eve does not have prior knowledge of the ISAC TX signal, as it is randomized by communication data.
    \item \textbf{(A.4)} The sensing Eve has no prior range-Doppler information of the targets.
    \item \textbf{(A.5)} The ISAC TX has no information regarding the sensing Eve, including Eve’s location.
\end{itemize}
The assumptions \textbf{(A.1)}-\textbf{(A.5)} are realistic in practical scenarios where the sensing Eve remains silent and unknown to the ISAC TX, posing a potential threat to ISAC sensing security. Notably, it should be emphasized that prior knowledge of Eve’s channel is not required in the proposed framework to ensure ISAC sensing security. This aligns with the worst-case security assumption, where the legitimate ISAC system operates without any prior information on Eve's location.

\subsection{Transmit Signal Model}
The ISAC TX signal, utilizing \(N\) OFDM subcarriers, is modulated with random communication symbols drawn from the constellation set \(\mathbb{S}\), which is expressed as $\mathbf{s} = [s_1, s_2, \dots, s_N]^T$, where \( s_n \in \mathcal{S}, \forall n = 1,2, \dots, N \). Without loss of generality, we assume that the constellations have zero-mean and unit-variance as $\mathbb{E}\left[|s_n|^2\right] = 1, \forall{s_n} \in \mathcal{S}$, of which statistical properties are defined as follows:
\begin{align}
    \mathbb{E}\left[|s_n|^4\right] = \mu_{4}, \;\;\; \mathbb{E}\left[|s_n|^{-2}\right]  = \nu_{-2}, \quad \forall{s_n} \in \mathcal{S}, \label{eq1}
\end{align}
where $\mu_{4}$ is the fourth moment of the constellation known as the kurtosis \cite{liu2024ofdm}, and $\nu_{-2}$ is the inverse second moment of the constellation \cite{wojaczek2018reciprocal}. For a general M-QAM constellation, they can be obtained by $\mu_{4} = \frac{1}{M}\sum_{m=1}^{M} |s_m|^4$ and $\nu_{-2} = \frac{1}{M}\sum_{m=1}^{M} |s_m|^{-2}$, respectively.

Furthermore, the power allocation of each individual subcarrier is embedded in the matrix \(\mathbf{W}\), given by $\mathbf{W} = \text{diag} (\mathbf{w})$, where $\mathbf{w} = \left[w_1, w_2, \dots, w_N \right]^T$ and $w_n$ denotes the power coefficient assigned to  subcarrier $n$. Accordingly, the frequency-domain representation of the ISAC TX signal is given by
\begin{equation}
    \mathbf{x} = \mathbf{W}\mathbf{s} \label{eq2}.
\end{equation}
The total transmit signal power per OFDM symbol encapsulated in \(\mathbf{W}\) is normalized to the unit-power, such that
\begin{align}
    \text{Tr}\left(\mathbf{W}\mathbf{W}^H\right) = N.
\end{align}
Taking the cyclic prefix (CP) into account, the CP-OFDM signal with a CP length of \( N_{\text{cp}} \) can be expressed in the frequency domain as \( \mathbf{x}_{\text{cp}} =[\mathbf{x}^T, \mathbf{0}^T_{N_{\text{cp}}}]^T \in \mathbb{C}^{(N + N_{\text{cp}}) \times 1} \). Since the CP is removed during receiver processing to mitigate inter-symbol interference (ISI) \cite{han2023sub}, it is omitted hereafter for clarity in the formulation. \textcolor{black}{In cases where the target delay exceeds the CP length, the useful signal power is reduced and ISI-induced interference arises, thereby degrading the sensing performance of CP-OFDM systems \cite{wang2023coherent}. Hence, in this work we assume that the delays of all targets-of-interest lie within the ISI-free region determined by the CP length, expressed as $R_{\text{max,cp}} = \frac{c N_{\text{cp}}}{2B}$, where $B$ denotes the OFDM signal bandwidth.}   

This unified ISAC signal is utilized for both communication and sensing. It is received by a communication user and is also reflected by targets, returning to both the legitimate sensing receiver and the sensing eavesdropper. Notably, the formulated ISAC signal in (\ref{eq2}) introduces a trade-off between S\&C performance. Specifically, the communication rate of the typical user and the range sidelobe level for sensing are influenced by the allocated subcarrier power \( w_1, w_2, \dots, w_N \). In the proposed framework for sensing security, our objective is to design ISAC signaling that optimally balances the three-way trade-offs between communication, sensing, and sensing security by carefully controlling the radar ambiguity through the power allocation across subcarriers.

\subsection{Communication System Model}
Let \(\mathbf{y}_c\) denote the received signal at the typical communication user. After removing the CP, the frequency-domain representation of the received signal is given by
\begin{align}
    \mathbf{y}_c = \mathbf{H}_c \mathbf{W} \mathbf{s} + \mathbf{z}_c,
\end{align}
where \(\mathbf{H}_c\) is a diagonal matrix representing the communication channel, given by \(\mathbf{H}_c = \text{diag}(h_1, h_2, \dots, h_N)\) with each entry $h_i$ representing the channel gain of the $i^{th}$ subcarrier. The term \(\mathbf{z}_c\) represents additive white Gaussian noise (AWGN), modeled as \(\mathbf{z}_c \sim \mathcal{CN}(0,\sigma_c^2 \mathbf{I}_N)\). Here, we consider a frequency-selective fading channel. However, each subband is assumed to be sufficiently narrow, such that the subband signals modulated on each subcarrier experience flat fading. Additionally, the fading characteristics of the channel remain constant over the duration of single transmission frame. Based on this model, the achievable communication rate at the typical user is given by
\begin{align}
    R_c = \frac{B}{N}\sum_{i=1}^N \log_2 \left(1+ \frac{|h_i|^2|w_i|^2}{\sigma_c^2} \right), \label{eq5}
\end{align}
where $B$ is the total bandwidth of the OFDM signal. \textcolor{black}{We adopt the achievable rate as the communication performance metric for ISAC signaling design, which is determined by the SNR of each subcarrier and, in turn, depends on the subcarrier power allocation.} 

\subsection{Sensing System Model: Legitimate Sensing Receiver (Alice)}
Given that \textcolor{black}{\( U = U_t + U_c \) radar reflections from $U_t$ targets and $U_c$ clutter sources} are located at different ranges, the received signal at the legitimate sensing receiver can be generally modeled as  
\begin{align}
    \mathbf{y}_{A,s} & = \textcolor{black}{\left(\sum_{i = 1}^{U_{t}}\beta_{A,i}^{(t)} \mathbf{r}\left(\tau_{A,i}^{(t)}\right) + \sum_{j = 1}^{U_{c}}\beta_{A,j}^{(c)} \mathbf{r}\left(\tau_{A,j}^{(c)}\right)\right)} \odot \mathbf{x} + \mathbf{z}_{A,s}, \nonumber \\
    & = \mathbf{h}_{A,s} \odot \mathbf{W} \mathbf{s} + \mathbf{z}_{A,s}. \label{eq7}
\end{align}
\textcolor{black}{Here, \( \beta_{A,i}^{(t)} \) denotes the complex amplitude that accounts for both the path loss and the radar cross-section (RCS) of target \( i \), while \( \tau_{A,i}^{(t)} \) represents the time-of-flight (TOF) from the ISAC transmitter to target \( i \) and back to Alice. Similarly, \( \beta_{A,j}^{(c)} \) and \( \tau_{A,j}^{(c)} \) denote the complex amplitude and TOF associated with clutter sources, respectively.} The range steering vector is given as  $\textbf{r}(\tau) = \begin{bmatrix} 1, e^{-j2 \pi \Delta f \tau}, \cdots, e^{-j2 \pi (N-1)\Delta f \tau} \end{bmatrix}^T \in \mathbb{C}^{N \times 1}$ with the subcarrier spacing \( \Delta f = B/N \). Thus, the \textcolor{black}{radar} channel vector of Alice $\mathbf{h}_{A,s}$ is given by \textcolor{black}{$\mathbf{h}_{A,s} = \left(\sum_{i = 1}^{U_{t}}\beta_{A,i}^{(t)} \mathbf{r}\left(\tau_{A,i}^{(t)}\right) + \sum_{j = 1}^{U_{c}}\beta_{A,j}^{(c)} \mathbf{r}\left(\tau_{A,j}^{(c)}\right)\right)$.}  The term \( \mathbf{z}_{A,s} \) represents the AWGN of the sensing receiver, following \( \mathbf{z}_{A,s} \sim \mathcal{CN}(0,\sigma_A^2 \mathbf{I}_N) \). We assume static or slowly moving targets, such that range migration and inter-carrier interference (ICI) effects are negligible. These effects, if present, would otherwise increase the overall sidelobe levels \cite{hakobyan2017novel, tigrek2010compensation}.

As a preliminary step to target detection and parameter estimation, we consider two types of radar receiver processing: matched filtering (MF) and reciprocal filtering (RF). The MF is well known as the optimal linear receiver filter in terms of maximizing the signal-to-noise ratio (SNR) at the target coordinates, without considering sidelobe effects. In contrast, RF is a form of mismatched filtering (MMF) that can be designed to reduce sidelobe levels at the expense of decreased SNR for the targets \cite{mercier2020comparison}.

\subsubsection{Matched filtering at Alice}
Let $\mathbf{g}_{A,MF}$ denote the frequency-domain MF of Alice, which is expressed as $\mathbf{g}_{A,MF} = \mathbf{x}^{*}$. Then, the output of the MF in the frequency-domain is given by
\begin{align}
    \mathbf{h}_{A,MF} & = \mathbf{y}_{A,s} \odot \mathbf{g}_{A,MF} \nonumber \\
    & = \mathbf{W}^2 \mathbf{S}^2\mathbf{h}_{A,s} + \mathbf{z}_{A,s} \odot \mathbf{x}^{*}, \label{eq8}
\end{align}
where $\mathbf{S} = \text{diag}(\mathbf{s})$. From the MF output, the range profile is obtained by performing inverse discrete Fourier transform (IDFT) over (\ref{eq8}). Let us define $\mathbf{F}_N$ as the normalized DFT matrix of size $N$. Then, the range profile $\mathbf{\Gamma}_{A,MF}$ is expressed as
\begin{align}
    \mathbf{\Gamma}_{A,MF} & = \mathbf{F}_N^H \mathbf{W}^2\mathbf{S}^2\mathbf{h}_{A,s} + \tilde{\mathbf{z}}_{A,MF,s}, \label{eq9}
\end{align}
where $\tilde{\mathbf{z}}_{A,MF,s}$ follows the noise statistics as $\tilde{\mathbf{z}}_{A,MF,s} \sim \mathcal{CN}(0,\sigma_A^2 \mathbf{I}_N)$.

\subsubsection{Reciprocal filtering at Alice}
Let $\mathbf{g}_{A,RF}$ denote the frequency-domain RF of Alice, which is expressed as $\mathbf{g}_{A,RF} = \mathbf{1}_{N} \oslash \mathbf{x}$. The output of the RF in the frequency-domain is given by
\begin{align}
    \mathbf{h}_{A,RF} & = \mathbf{y}_{A,s} \odot \mathbf{g}_{A,RF} \nonumber \\
    & = \mathbf{h}_{A,s} + \mathbf{z}_{A,s} \oslash \mathbf{x}. \label{eq10}
\end{align}
Performing IDFT over the RF output, the range profile with the RF can be obtained as
\begin{align}
    \mathbf{\Gamma}_{A,RF} & = \mathbf{F}_N^H \mathbf{h}_{A,s} + \tilde{\mathbf{z}}_{A,RF,s}, \label{eq11}
\end{align}
where $\tilde{\mathbf{z}}_{A,RF,s} \sim \mathcal{CN}(0, \frac{1}{N}\sum_{n=1}^N |w_n|^{-2} \nu_{-2} \sigma_A^2 \mathbf{I}_N)$.

\textbf{Remark 1:}  Inspired by the characteristics of these receivers, we aim to design the ISAC signaling to enhance sensing security while ensuring favorable sensing performance at the legitimate sensing receiver. The key idea for the legitimate sensing receiver is to exploit the RF for mitigating the undesired sidelobes caused by the design of the secure ISAC signal. However, the use of the RF in lieu of MF at Alice inherently causes the degradation on the output SNR. Importantly, this introduces the trade-off between the legitimate sensing performance and the sensing security, of which metrics are detailed in Section \ref{metric_Alice} and \ref{metric_Eve}. 

\subsection{Sensing System Model: Sensing Eavesdropper (Eve)}
The sensing eavesdropper, which is a passive bistatic sensing receiver without prior knowledge of the ISAC TX signals, requires a reference signal to demodulate the surveillance signal from targets. Based on assumption (\textbf{A.1}), we model the surveillance signal and the reference signal separately. Similar to (\ref{eq7}), the received signal at Eve, \textcolor{black}{reflected from \( U = U_t + U_c \) sources including multiple targets and clutter}, is given by  
\begin{align}
    \mathbf{y}_{E,s} & = \textcolor{black}{\left(\sum_{i = 1}^{U_{t}}\beta_{E,i}^{(t)} \mathbf{r}\left(\tau_{E,i}^{(t)}\right) + \sum_{j = 1}^{U_{c}}\beta_{E,j}^{(c)} \mathbf{r}\left(\tau_{E,j}^{(c)}\right)\right)} \odot \mathbf{x} + \mathbf{z}_{E,s} \nonumber \\
    & = \mathbf{h}_{E,s} \odot \mathbf{W} \mathbf{s} + \mathbf{z}_{E,s},
\end{align}
\textcolor{black}{where \( \beta_{E,i}^{(t)} \) and \( \beta_{E,j}^{(c)} \) are the complex amplitudes of targets and clutter sources, respectively, and \( \tau_{E,i}^{(t)} \) and \( \tau_{E,j}^{(c)} \) are the TOFs from the ISAC TX to reflection source and then to Eve.} The term \( \mathbf{z}_{E,s} \) represents the AWGN at Eve's receiver, following \( \mathbf{z}_{E,s} \sim \mathcal{CN}(0,\sigma_{E}^2 \mathbf{I}_N) \).

The reference signal at Eve, which is leaked from the ISAC TX through the surveillance link in Fig. \ref{f1}, is modeled as
\begin{align}
    \mathbf{y}_{E,r} & = \mathbf{h}_{E,r} \odot \mathbf{x} + \mathbf{z}_{E,r}. \label{eq13}
\end{align}
Here, we assume that Eve's channel \( \mathbf{h}_{E,r} \) follows the Rician channel model, given by \cite{panayirci2019sparse}
\begin{align}
    \mathbf{h}_{E,r} & = \sqrt{\frac{g_{E,r} K}{K+1}} \mathbf{h}_{E,LoS} + \sqrt{\frac{g_{E,r}}{K+1}} \mathbf{h}_{E,NLoS}, \label{eq14}
\end{align}
where \( \mathbf{h}_{E,LoS} \) represents the line-of-sight (LoS) component with a normalized channel gain, and \( \mathbf{h}_{E,NLoS} \) represents the non-line-of-sight (NLoS) component, which follows a complex Gaussian distribution with zero mean and unit variance. The term \( g_{E,r} \) denotes Eve's channel gain, while \( K \) determines the relative power between the LoS and NLoS components. Additionally, assumption (\textbf{A.2}) allows Eve to have prior knowledge of \( \mathbf{h}_{E,LoS} \). By substituting (\ref{eq14}) into (\ref{eq13}) and removing \( \mathbf{h}_{E,LoS} \), the reference signal at Eve can be rewritten as
\begin{align}
    \tilde{\mathbf{y}}_{E,r} & = \sqrt{\frac{g_{E,r} K}{K+1}}\mathbf{x} + \sqrt{\frac{g_{E,r}}{K+1}} \tilde{\mathbf{h}}_{E,NLoS} \odot \mathbf{x} + \tilde{\mathbf{z}}_{E,r}, \label{eq15}
\end{align}
where \( \tilde{\mathbf{h}}_{E,NLoS} = \mathbf{h}_{E,NLoS} \odot \mathbf{h}_{E,LoS}^* \) and \( \tilde{\mathbf{z}}_{E,r} \) share the same statistical properties as \( \mathbf{h}_{E,NLoS} \) and \( \mathbf{z}_{E,r} \), respectively.

\begin{figure}[t!]
    \centering
    {\includegraphics[scale=0.75]{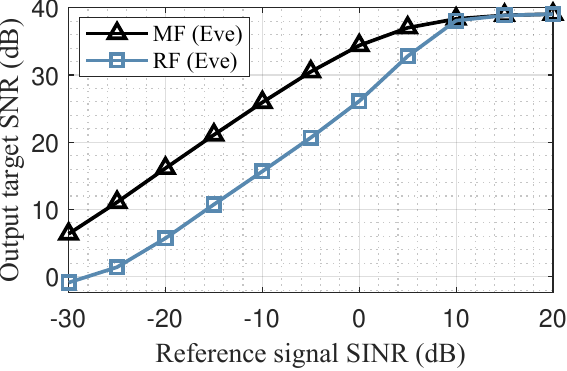}} 
    \caption{The target SNR at the filter output vs. Eve's reference signal SINR with QPSK constellation and equally-allocated subcarrier power. The input target SNR is 0 dB, and 32 OFDM symbols with $N = 256$ are coherently processed, \textcolor{black}{yielding 39 dB of the processing gain for a static target.}}
    \label{f2}
\end{figure}
\textbf{Remark 2:} Importantly, unlike the legitimate sensing receiver, Eve must exploit the reference signal (\ref{eq15}) to estimate the sensing channel\cite{cui2014target}. However, this reference signal is affected by receiver noise as well as multi-path reflection components, which degrade the signal-to-interference-plus-noise ratio (SINR) \cite{bkaczyk2015impact}. Accordingly, performing RF at Eve is unfavorable, as it further degrades the target SNR, leading to poor target detection and estimation performance. Fig. \ref{f2} emphasizes the output SNR gap between MF and RF at Eve along the reference signal SINR. Thus, to avoid severe SNR degradation, Eve leverages MF to extract the sensing channel from the surveillance signal. \textcolor{black}{Additionally, performing mismatched filtering at the sensing receiver requires full knowledge of the transmitted signal, which prevents Eve from employing advanced receiver processing without access to this information in practice. In summary, Table \ref{t1} outlines the operational differences between Alice and Eve that are leveraged in the design of sensing-secure ISAC signaling.}

\arrayrulecolor{black}
\begin{table}[t!]
\centering
    \fontsize{8.3}{14}\selectfont
    \caption{\textcolor{black}{Operational Differences between Alice and Eve utilized for ISAC sensing security}} 
        \label{t1}
    \begin{tabular}{>{\centering\arraybackslash} m{5em} | >{\centering\arraybackslash} m{4.5em}  | >
    {\centering\arraybackslash} m{5em}  | >
    {\centering\arraybackslash} m{5em}  | >{\centering\arraybackslash} m{4em}}
        \toprule
            & \textcolor{black}{Sensing mode} & \textcolor{black}{TX signal knowledge} & \textcolor{black}{Receiver processing} &
            \textcolor{black}{Resulting effects}\\ 
            \hline
        \midrule
        \textcolor{black}{Legitimate Sensing (Alice)}  & \textcolor{black}{Monostatic or bistatic (Active)} & \textcolor{black}{Fully-known} & \textcolor{black}{Reciprocal filtering } & \textcolor{black}{SNR loss}       \\ \hline
        \textcolor{black}{Eavesdropper (Eve)}   & \textcolor{black}{Bistatic (Passive)} & \textcolor{black}{Unknown} & \textcolor{black}{Matched filtering}  & \textcolor{black}{Sidelobes}      \\ 
        \bottomrule
    \end{tabular}
\end{table}

For the MF at Eve, let us denote \( \mathbf{g}_{E,MF} = \tilde{\mathbf{y}}_{E,r}^* \) as the frequency-domain MF of Eve. The MF output is given by
\begin{align}
    \mathbf{h}_{E,MF}  = & \sqrt{\frac{g_{E,r} K}{K+1}} \mathbf{W}^2 \mathbf{S}^2 \mathbf{h}_{E,s} + \sqrt{\frac{g_{E,r}}{K+1}} \mathbf{W}^2 \mathbf{S}^2\mathbf{h}_{E,s} \odot \tilde{\mathbf{h}}_{E,NLoS}^* \nonumber \\
    & \qquad \qquad \qquad \quad   + \mathbf{y}_{E,s} \odot \tilde{\mathbf{z}}_{E,r}^* + \mathbf{z}_{E,s} \odot {\mathbf{y}}_{E,r}^*. \label{eq16}
\end{align}
In (\ref{eq16}), it is readily observed that the MF output at Eve consists of the \textcolor{black}{radar} channel and additive noise terms, which are determined by the SINR of the reference signal and the noise characteristics of Eve’s receiver. Since these additive noise components cannot be controlled by the ISAC TX, securing the sensing functionality must instead rely on the ambiguity function (AF) of the ISAC signal, which directly impacts Eve's MF output. \textcolor{black}{For the MF receiver, a high peak sidelobe level (PSL) and integrated sidelobe level (ISL) in the AF increase the risk of false detections, requiring the radar system to use higher detection thresholds to maintain acceptable false alarm rates. Moreover, elevated ISL degrades the accuracy of parameter estimation in multi-target scenarios \cite{altes1979target, de2020sidelobe}. In the context of sensing security, where Eve is constrained to matched filtering due to limited knowledge of the transmitted signal, high PSL and ISL values significantly hinder both detection and estimation performance.} Based on this observation, we develop a framework for sensing-secure ISAC signaling by controlling the radar sensing ambiguity, as detailed in the following sections.

\section{Ambiguity Function Engineering for Sensing Security in ISAC: Artificial Target Generation}
Our key idea for achieving secure sensing in ISAC is to generate artificial targets for the unknown sensing eavesdropper while allowing the legitimate sensing receiver to eliminate them using the RF based on the known ISAC signal. This is accomplished by designing the AF of the ISAC signal to exhibit multiple ambiguous peaks with magnitudes comparable to that of the AF's mainlobe. Here, we focus on the artificial targets in the range profile of Eve. Therefore, we primarily investigate the auto-correlation function (ACF) of the ISAC signal, which corresponds to the zero-Doppler cut of the AF. \textcolor{black}{It is important to note that the proposed framework emphasizes the AF characteristics of ISAC signaling, which can lead to false detections and large-scale estimation errors at Eve, rather than relying on estimation-theoretic approaches that primarily address small-scale, unbiased estimation performance.}

\subsection{Ambiguity Function Shaping for Artificial Targets}
\begin{figure}[t!]
    \centering
    \subfloat[]{\includegraphics[scale=0.9]{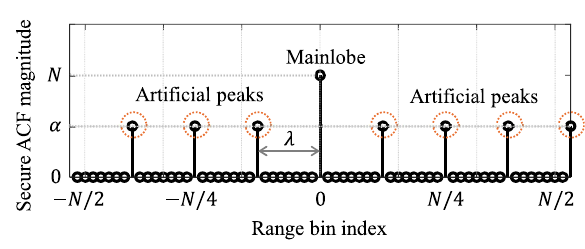}} \\
    \centering
    \subfloat[]{\includegraphics[scale=0.9]{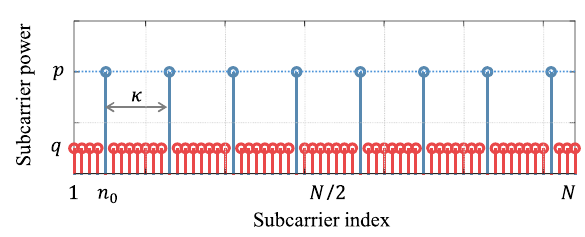}} 
    \caption{(a) The secure ACF with artificial peaks to achieve secure sensing in ISAC, and (b) the subcarrier power allocation scheme corresponding to the secure ACF.}
    \label{f3}
\end{figure}
As Eve's range profile with the MF is only controllable through the ACF of the ISAC signal, we define the frequency-domain ACF as
\begin{align}
    \mathbf{\Lambda} = \sqrt{N}\mathbf{F}_N^H \mathbf{W}^2 \mathbf{S}^2 \mathbf{1}_{N},  
\end{align}
where the \( k \)th element is expanded in scalar form as
\begin{align}
    \Lambda[k] = \sum_{n=1}^N |w_n|^2|s_n|^2e^{j\frac{2\pi}{N}k(n-1)}, \quad \forall k. \label{eq18} 
\end{align}
It is noteworthy that the ACF exhibits an impulse function if equal power allocation is applied, i.e., \( w_1 = w_2 = \dots = w_N \), and the unit-amplitude constellation with $|s_n|=1, \forall{n}$ is employed. This is in line with the understanding that PSK is optimal for OFDM sensing as proven in \cite{liu2024ofdm}. The proposed approach for securing sensing functionality is to generate deterministic periodic peaks in the ACF. Since Eve lacks prior information about the actual targets, these periodic peaks act as artificial targets, misleading Eve's target estimation.

Let us define \( \tilde{\Lambda}[k] \) as the ideal ACF for the artificial target generation. Then, the desired ACF for secure sensing (denoted as the \textit{"secure ACF"}) can be represented as a Dirac comb function, with its initial formulation given by  
\begin{align}
    \tilde{\Lambda}[k] = \underbrace{N\delta[k]}_{\text{Mainlobe}} + \underbrace{\sum_{l=1}^{L} \alpha \delta[k-l\lambda]}_{\text{Artificial peaks}}, \quad \forall k.  \label{eq19}  
\end{align}
Here, \( \lambda \) represents the periodicity of the peaks, \( L \) denotes the number of artificial peaks in the ACF, given by \( L = N/\lambda - 1 \), and \( \alpha \) represents the artificial peak magnitude. This secure ACF is illustrated in Fig. \ref{f3}(a). More explicitly, it is well known that the maximum unambiguous range of OFDM radar is given by \( R_{\max} = \frac{cN}{2B} \), where \( c \) is the speed of light. The artificial peaks, relative to the zero-delay position, are equivalently located at ranges \( \left[\frac{cN}{2B(L+1)}, \frac{2cN}{2B(L+1)}, \dots, \frac{cNL}{2B(L+1)}\right] \), which limits the unambiguous range to \( \frac{cN}{2B(L+1)} \). Notably, sensing security improves when the secure ACF contains a greater number of artificial peaks and higher artificial peak magnitudes, as these are perceived as potential artificial targets by Eve.

\textcolor{black}{\textbf{Remark 3:}} \textcolor{black}{While our primary focus is on shaping the AF in the range domain through subcarrier power allocation in OFDM, the proposed approach can be naturally extended to enhance velocity-domain sensing security via symbol-by-symbol power allocation across the slow-time domain. This extension effectively shapes the zero-delay cut of the AF. Specifically, the maximum unambiguous Doppler frequency in OFDM radar with a symbol repetition interval of \( T_{\text{sym}} \) is given by \( f_{d,\max} = \frac{1}{2T_{\text{sym}}} \). By applying OFDM symbol power allocation, $L$ number of artificial peaks can be introduced in the Doppler domain at frequency bins \(\left[\frac{1}{2T_{\text{sym}}(L+1)}, \frac{2}{2T_{\text{sym}}(L+1)}, \dots, \frac{L}{2T_{\text{sym}}(L+1)}\right],\) thereby reducing the effective unambiguous Doppler to \( \frac{1}{2T_{\text{sym}}(L+1)} \). For clarity of exposition, this paper remains focused on range-domain analysis.}

The secure ACF with deterministic artificial peaks can be realized through the design of subcarrier power allocation \( \mathbf{W} \). Provided that the modulated symbol comes from a unit-amplitude constellation, such as PSK, the allocated subcarrier power for achieving the secure ACF is given by the following theorem. 
\begin{theorem}\label{theo1}
    The power allocation of subcarriers for the secure ACF in (\ref{eq19}) is given by
    \begin{align}
        |w_n|^2 = \begin{cases}
                p, & \text{if } n \in \mathcal{P}_{1,\kappa}=\left\{1,1+\kappa, 1+ 2\kappa , \dots \right\} \\
                q, & \text{if } n \in \mathcal{P}_{1,\kappa}^c
                \end{cases}, \label{eq20}
    \end{align}
    where \( p > q > 0 \), $\kappa$ is a divisor of $N$, $p + (\kappa - 1)q = \kappa$, and
    \begin{align}
        (p,q,\kappa)  = \left(1 +\frac{\alpha L}{N}, 1-\frac{\alpha}{N}, L+1 \right). \label{eq20-2}
    \end{align}
    The set \( \mathcal{P}_{1,\kappa} \) represents the uniformly spaced subcarriers, starting from the first subcarrier and spaced by \( \kappa \), which have the dominant power \( p \).
\end{theorem}
\renewcommand\qedsymbol{$\blacksquare$}
\begin{proof}
Please refer to Appendix \ref{proof_theo1}.
\end{proof}
The above theorem establishes the foundation for structured subcarrier power allocation to achieve a secure ACF. However, it only provides a fixed subcarrier power allocation for sensing-secure ISAC signaling, which lacks sufficient degrees of freedom (DoFs) to further optimize the balance between S\&C performance. To address this limitation, we verify the following corollaries for advanced secure-sensing ISAC signaling, incorporating deterministic artificial peaks in the ACF to enable greater flexibility in signaling design.

\begin{corollary}\label{corr1}
    (Shifted-Invariant Property) The circularly shifted subcarrier power allocation of (\ref{eq20}), expressed as
    \begin{align}
        |w_n|^2 = \begin{cases}
                p, & \text{if } n \in \mathcal{P}_{n_0,\kappa}=\left\{n_0,n_0+\kappa, n_0+ 2\kappa , \dots \right\} \\
                q, & \text{if } n \in \mathcal{P}_{n_0,\kappa}^c
                \end{cases}, 
    \end{align}
    where \( n_0 \leq \kappa \), results in a squared ACF identical to the squared secure ACF of (\ref{eq19}), i.e., \( \left|\tilde{\Lambda}[k]\right|^2 \).
\end{corollary}
\renewcommand\qedsymbol{$\blacksquare$}
\begin{proof}
Please refer to Appendix \ref{proof_corr1}.
\end{proof}

\begin{corollary}\label{corr2}
    The subcarrier power allocation following the distribution
    \begin{align}
        \mathbb{E}\left[|w_n|^2\right] = \begin{cases}
                p, & \text{if } n \in \mathcal{P}_{n_0,\kappa}=\left\{n_0,n_0+\kappa , \dots \right\} \\
                q, & \text{if } n \in \mathcal{P}_{n_0,\kappa}^c
                \end{cases}, \label{eq22}
    \end{align}
    approximately results in an expected squared ACF identical to the squared secure ACF of (\ref{eq19}), i.e., \( \left|\tilde{\Lambda}[k]\right|^2 \), provided that \( \text{Var}\left[|w_n|^2\right] \) is sufficiently small.
\end{corollary}
\renewcommand\qedsymbol{$\blacksquare$}
\begin{proof}
Please refer to Appendix \ref{proof_corr2}.
\end{proof}
\noindent The structured power allocation scheme for the secure ACF is graphically illustrated in Fig. \ref{f3}(b). Consequently, the power allocation of subcarriers enables the intentional generation of deterministic artificial peaks in the ACF, where the number and magnitude of these peaks are determined by the allocation parameters \( (p, q, \kappa) \) given in (\ref{eq20-2}).

Now, we consider the random ISAC signals based on the general constellation set, which indeed have the same artificial peaks in the ACF under the uniformly sparse subcarrier power allocation. In this regard, we evaluate the expectation of the squared ACF based on (\ref{eq20}). This is given by a closed form as follows:

\begin{proposition}\label{prop1}
    The expectation of the squared secure ACF under the subcarrier power allocation (\ref{eq20}) and the constellation set with the kurtosis $\mu_{4}$ is expressed as
    \begin{align}
    \mathbb{E}\left[|{\Lambda}[k]|^2\right] = & \underbrace{N^2{\delta[k]}}_{\text{Mainlobe}} + \underbrace{\alpha^2 \sum_{l=1}^{L} \delta[k-l\lambda]}_{\text{Artificial peaks}}  \nonumber\\
    & +  \underbrace{(\mu_{4}-1) \left(\frac{N}{\kappa}p^2 + N\left(1 - \frac{1}{\kappa}\right)q^2 \right)}_{\text{Sidelobe caused by random signaling}}. \label{eq23}
    \end{align}
\end{proposition}
\renewcommand\qedsymbol{$\blacksquare$}
\begin{proof}
Please refer to Appendix \ref{proof_prop1}.
\end{proof}

\textbf{Remark 4:} Notably, the ACF presented in (\ref{eq19}) can be achieved only under unit-amplitude constellations with the kurtosis $\mu_{4} = 1$. This is because other constellations with kurtosis \( u_4 > 1 \) inherently introduce additional sidelobes across all range bins in the ACF, as observed in (\ref{eq23}) and corroborated by \cite{liu2024ofdm, liu2025uncovering}. Nevertheless, the proposed secure ACF with deterministic artificial peaks under the general constellation as QAM can still be achieved using the subcarrier power allocation scheme (\ref{eq20}) as presented in Proposition \ref{prop1}, while including additional sidelobes induced by the constellation. Indeed, these additional sidelobes are useful to enhance the sensing security.

\subsection{Performance of Legitimate Sensing Receiver (Alice)} \label{metric_Alice}
In the proposed framework for secure sensing ISAC, the legitimate receiver exploits RF to cancel out the artificial peaks generated through ambiguity shaping as provided in \textcolor{black}{\eqref{eq10}}. \textcolor{black}{This is because the RF operation equalizes the effect of transmitted signals on the \textcolor{black}{radar sensing} channels \cite{mercier2020comparison}}. Although the output of RF is free from the effects of range sidelobes caused by the secure ACF, RF suffers from noise enhancement, leading to a decrease in SNR when the noise is divided by a subcarrier element with very small power allocation. Accordingly, we introduce the SNR loss of RF relative to the SNR of MF as the performance metric of Alice, which is defined as follows \cite{rodriguez2023supervised}:
\begin{align}
    \mathcal{L}_{A} =  \frac{\gamma_{MF}}{\gamma_{RF}}, \label{eq24}
\end{align}
where \( \gamma_{MF} \) and \( \gamma_{RF} \) denote the SNRs of a typical target at the MF and RF outputs, respectively. This SNR loss indicates the amount of the noise amplification by RF compared to that of MF, which is the optimal receiver in terms of the SNR.  

Given a scalar form of a range profile $\Gamma[n]$ and a target at the range bin of $n_t$, the output SNR is explicitly written as
\begin{align}
    \gamma =  \frac{\left|\mathbb{E}\left[\Gamma[n_t]\right] \right|^2}{\mathbb{E}\left[|\Gamma[n]|^2\right]}. \label{eqSNR}
\end{align}
With (\ref{eq9}), (\ref{eq11}), and (\ref{eqSNR}) at hand, we derive the output SNR of the respective receiver at Alice, which are given by the following lemma.
\begin{lemma}\label{lemma1}
    The output SNRs of the MF and RF at Alice's receiver are respectively given by
    \begin{align}
        \gamma_{MF} & =  N \beta_{A}^2\sigma_{A}^{-2}, \label{eq25} \\
        \gamma_{RF} & =  N \beta_{A}^2\sigma_{A}^{-2} \left( \frac{N}{\nu_{-2}\sum_{n=1}^N|w_n|^{-2}}\right). \label{eq26}
    \end{align}
\end{lemma}
\renewcommand\qedsymbol{$\blacksquare$}
\begin{proof}
Please refer to Appendix \ref{proof_lemma1}.
\end{proof}
Now, we are ready to explicitly express the SNR loss at Alice as a function of the ISAC TX signal.
\begin{theorem}\label{theo2}
    The SNR loss of the RF output relative to the SNR of the MF output is given by
    \begin{align}
        \mathcal{L}_{A} =  \frac{\nu_{-2}}{N} \sum_{n=1}^N |w_n|^{-2}. \label{eqtheo1}
    \end{align}
\end{theorem}
\renewcommand\qedsymbol{$\blacksquare$}
\begin{proof}
Substituting (\ref{eq25}) and (\ref{eq26}) into (\ref{eq24}) directly yields the result.
\end{proof}

\begin{corollary}\label{corr3}
    The SNR loss with the subcarrier power allocation (\ref{eq20}) for the secure ACF is given by
    \begin{align}
        \mathcal{L}_A =  \frac{\nu_{-2}}{\kappa} \left(\frac{1}{p} + \frac{\kappa-1}{q} \right). \label{eqCorr3}
    \end{align}
\end{corollary}
\renewcommand\qedsymbol{$\blacksquare$}
\begin{proof}
Substituting (\ref{eq20}) into (\ref{eqtheo1}) yields (\ref{eqCorr3}).
\end{proof}

\textbf{Remark 5:} Clearly, the defined SNR loss depends solely on the ISAC TX signal, specifically its constellation and subcarrier power allocation. The minimum SNR loss is 0 dB, which is achieved under a unit-amplitude constellation and an equal power allocation scheme. From (\ref{eqCorr3}), it is intuitively observed that higher artificial peaks in the secure ACF result in greater SNR loss, implying that more secure signaling degrades legitimate sensing performance. This reveals an inherent performance trade-off between legitimate sensing and sensing security in the proposed framework.

\subsection{Performance of Sensing Eavesdropper (Eve)} \label{metric_Eve}
To assess the sensing security of ISAC signaling, we leverage \textcolor{black}{PSL and ISL} as sensing security metrics. Since Eve's receiver performance cannot be directly evaluated, PSL and ISL of the ACF serve as proxies for measuring the sensing security level. \textcolor{black}{Moreover, the PSL and ISL of the AF determine the sensing performance of the MF, which is directly linked to Eve's sensing performance.} A high PSL increases the likelihood of false target detection by Eve, as it corresponds to the magnitude of artificial targets generated through ambiguity shaping. Additionally, a high ISL leads to erroneous range estimation by Eve and may prevent the detection of weak targets when multiple targets exist. In our framework, ISL is influenced by the number of artificial targets and sidelobe levels caused by random signaling. Notably, these metrics represent the worst-case security level, as Eve's receiver output is further degraded due to errors in the estimation of the reference probing signal.

\subsubsection{Peak Sidelobe Level}
Let us define PSL as the magnitude of the highest sidelobe level relative to that of the mainlobe. This is expressed as  
\begin{align}
    \Delta_{\text{PSL,E}} = \frac{\max_{k \neq 0}\mathbb{E}\left[|{\Lambda}[k]|^2\right]}{\mathbb{E}\left[|{\Lambda}[0]|^2\right]}. 
\end{align}
Under signaling with the secure ACF, it is immediately expressed in the following theorem.  
\begin{theorem}\label{theo3}
    Under the subcarrier power allocation with \( (p,q,\kappa) \) in (\ref{eq20}), the PSL of the squared secure ACF is approximately given by
    \begin{align}
        \Delta_{\text{PSL,E}} = (1-q)^2. \label{eq27}
    \end{align}
\end{theorem}
\renewcommand\qedsymbol{$\blacksquare$}
\begin{proof}
From (\ref{eq23}), the magnitude of the mainlobe level approximates \( N^2 \) for sufficiently large \( N \). Additionally, the highest sidelobe level equals the magnitude of artificial peaks, denoted as \( \alpha^2 \). Based on \( \alpha = N(1-q) \) from (\ref{eq20-2}), the ratio \( \alpha^2/N^2 \) simplifies to \( (1-q)^2 \), completing the proof.
\end{proof}

Here, sparse power allocation with \( q=0 \) generates artificial targets with a magnitude equal to that of the mainlobe. This implies that such a subcarrier power allocation results in a PSL of 0 dB, causing the artificial targets to have the same magnitude as the true targets in Eve's receiver. However, it is important to note that the case of \( q=0 \) is not considered, as Alice would be unable to perform RF due to division by zero. Additionally, this condition ensures that communication data is always transmitted across all subcarriers.

\subsubsection{Integrated Sidelobe Level}
The ISL is defined as the total power contained in all sidelobes relative to the power in the mainlobe, which is given by 
\begin{align}
    \Delta_{\text{ISL,E}} = \frac{\sum_{k \neq 0}^N\mathbb{E}\left[|{\Lambda}[k]|^2\right]}{\mathbb{E}\left[|{\Lambda}[0]|^2\right]}. 
\end{align}
Then, the following theorem gives the ISL under the secure sensing signaling.
\begin{theorem}\label{theo4}
    Under the subcarrier power allocation with \( (p,q,\kappa) \) in (\ref{eq20}), the ISL of the squared secure ACF is approximately given by
    \begin{align}
        \Delta_{\text{ISL,E}} = (\kappa-1)(1-q)^2 + (\mu_{4}-1) \left(\frac{p^2}{\kappa} + \left(1 - \frac{1}{\kappa}\right)q^2\right). \label{eq29}
    \end{align}
\end{theorem}
\renewcommand\qedsymbol{$\blacksquare$}
\begin{proof}
From (\ref{eq23}), the total power in sidelobes is computed as
\begin{align}
    & \sum_{k \neq 0}^N\mathbb{E}\left[|{\Lambda}[k]|^2\right]  = L\alpha^2 + N(N-1)(\mu_{4}-1) \left(\frac{p^2}{\kappa} + \left(1 - \frac{1}{\kappa}\right)q^2 \right) \nonumber \\
    & \approx N^2\left((\kappa-1)(1-q)^2 + (\mu_{4}-1) \left(\frac{p^2}{\kappa} + \left(1 - \frac{1}{\kappa}\right)q^2 \right)\right).
\end{align}
For sufficiently large $N$, the mainlobe magnitude is given as $N^2$. Thus, this completes the proof.
\end{proof}

\begin{corollary}\label{corr4}
    The ISL of the secure ACF has a linear relationship with the PSL, expressed as  
    \begin{align}
        \Delta_{\text{ISL,E}} = \mu_{4}(\kappa-1)\Delta_{\text{PSL,E}} + (\mu_{4}-1). \label{eq30}
    \end{align}
\end{corollary}
\renewcommand\qedsymbol{$\blacksquare$}
\begin{proof}
From (\ref{eq27}), we have \( q = 1 - \sqrt{\Delta_{\text{PSL,E}}} \). Combining this with \( p + \left(\kappa - 1\right)q = \kappa \) gives \( p = (\kappa-1)\sqrt{\Delta_{\text{PSL,E}}} +1 \). Substituting \( p \) and \( q \) into (\ref{eq29}) yields (\ref{eq30}), completing the proof.
\end{proof}
From (\ref{eq30}), we observe that the ISL is influenced by three factors: the constellation, the number of artificial peaks, and their magnitude relative to the mainlobe. In this framework, we primarily focus on subcarrier power allocation to design sensing-secure ISAC signaling with deterministic artificial peaks, without delving deeply into the constellation set.

\section{Signaling Design for Sensing-Secure ISAC Based on AF Shaping} \label{section4}
Building on the above AF analysis, in this section, we present a sensing-secure ISAC signaling design with artificial targets, offering three-way trade-offs between communication, legitimate sensing, and sensing security. Notably, the subcarrier power allocation presented for the secure ACF in Theorem \ref{theo1} appears to provide a trade-off only between ISAC performance and sensing security. This is because it fixes the power of subcarriers in \( \mathcal{P}_{n_0,\kappa} \) and \( \mathcal{P}^c_{n_0,\kappa} \) to the exact values \( p \) and \( q \), respectively, without any degree of freedom (DoF) for ISAC signaling design under a given sensing security level. Nevertheless, Corollary \ref{corr2} introduces a certain level of DoF, enabling a more flexible trade-off between S\&C performance. Accordingly, we first investigate the trade-off between ISAC performance and sensing security with artificial targets and then explore a sensing-secure ISAC signaling design under a given level of sensing security.
     
\subsection{Trade-off between ISAC Performance and Sensing Security}
Firstly, we analyze the trade-off between ISAC performance and sensing security in the proposed framework. To this end, we determine a feasible performance region for secure-sensing ISAC based on Theorem \ref{theo1}, where communication, sensing, and sensing security performance are directly influenced by the selection of parameters \( (p,q,\kappa) \) for the secure ACF with artificial peaks. Recalling that $N$ subcarriers in OFDM signal are modulated by the constellation set $\mathcal{S}$, let us define \( \mathcal{C}(N,\mathcal{S}) \) as the set of feasible performance points, given by
\begin{align}
    & \mathcal{C}(N,\mathcal{S}) = \nonumber \\
    & \left\{(R_c, \mathcal{L}_A, \Delta_{\text{PSL,E}},  \Delta_{\text{ISL,E}}) \; | \; \text{(\ref{eq5}), (\ref{eqCorr3}), (\ref{eq27}), and (\ref{eq29})}, \forall{p,q,\kappa} \right\}, \label{eq35}
\end{align}
where \( p, q \), and \( \kappa \) are constrained as per Theorem \ref{theo1}. All operating points of \( \mathcal{C}(N,\mathcal{S}) \) can be determined by evaluating all possible combinations of \( (p,q,\kappa) \). Assuming a flat-fading communication channel, there is no clear trade-off between S\&C performance under the fixed power allocation scheme in Theorem \ref{theo1}. This is because the power allocation parameters \( (p,q,\kappa) \) that reduce the SNR loss of the legitimate sensing receiver also increase the communication rate in (\ref{eq5}). However, a flexible power allocation scheme based on Corollary \ref{corr2} enables a trade-off design between S\&C performance under a frequency-selective fading channel, which will be detailed in Section \ref{signaling}.

Instead, the performance trade-off can be observed between ISAC and sensing security by varying the parameters \( (p,q,\kappa) \) in (\ref{eq35}). It is straightforward to verify from (\ref{eq27}) and (\ref{eq30}) that higher values of \( \Delta_{\text{PSL,E}} \) and \( \Delta_{\text{ISL,E}} \) require a smaller \( q \) and a larger \( \kappa \). However, this increases the SNR loss, as shown in (\ref{eqCorr3}), thereby degrading both the legitimate sensing performance and the communication performance. The legitimate sensing optimal point in (\ref{eq35}) is achieved with equal subcarrier power allocation, i.e., \( \kappa = 1 \), resulting in the following metrics:
\begin{align}
    \mathcal{L}_{A}^{\text{sen-opt}} & =  \nu_{-2}, \\
    \Delta_{\text{PSL,E}}^{\text{sen-opt}} & = 0, \quad \Delta_{\text{ISL,E}}^{\text{sen-opt}} = \mu_{4}-1.
\end{align}
On the other hand, the sensing security optimal point in (\ref{eq35}) is obtained when \( \kappa = N/2 \) and \( q \rightarrow 0 \), yielding
\begin{align}
    \mathcal{L}_{A}^{\text{sec-opt}} & =  \infty, \\
    \Delta_{\text{PSL,E}}^{\text{sec-opt}} & = 1, \quad \Delta_{\text{ISL,E}}^{\text{sec-opt}} = \frac{\mu_{4}N}{2} - 1.
\end{align}
Accordingly, one may coarsely determine an appropriate sensing security level to ensure a desired level of ISAC performance, or vice versa, based on (\ref{eq35}).

\subsection{Sensing-Secure ISAC Signaling Design} \label{signaling}
We further investigate the optimization of sensing-secure ISAC signaling design with artificial targets, which provides a flexible balance between S\&C performance under a given level of sensing security. Based on Corollary \ref{corr2}, we can finely tune the subcarrier power allocation to optimize the S\&C performance bound while ensuring the desired level of sensing secrecy. Clearly, given that channel state information (CSI) is available at the ISAC TX, we can finely adjust \( \{w_n\}, \forall n \), to balance the SNR loss at Alice and the communication rate.

To this end, we formulate the following sensing-secure ISAC signaling design problem:
\begin{subequations}\label{Eqn::P1}
    \begin{align}
    (\mathbf{P.1}) \; & \underset{\{w_n\}}{\text{maximize}}
    \qquad   -(1-\rho) \frac{\mathcal{L}_{A}}{\mathcal{L}_{A,\rho=0}} + \rho \frac{R_c}{R_{c,\rho=1}} \label{Eqn::P1-0a} \\
    & \text{subject to}
    \qquad \qquad \qquad   \Delta_{\text{ISL,E}} \geq \epsilon_{\text{ISL}} ,   \label{Eqn::P1-0b} \\
    &  \qquad \qquad \qquad \qquad \qquad  \Delta_{\text{PSL,E}} \geq \epsilon_{\text{PSL}}, \label{Eqn::P1-0c}\\ 
    &  \qquad \qquad \qquad \qquad \qquad  \sum_{n=1}^N |w_n|^2 = N, \label{Eqn::P1-0d}
    \end{align}
\end{subequations} 
where \( \rho \in [0,1] \) is a weighting factor that determines the priority between sensing and communication functionalities. The S\&C performance metrics in the objective function (\ref{Eqn::P1-0a}) are normalized by their respective values obtained when \( \rho = 0 \) and \( \rho = 1 \). The parameters \( \epsilon_{\text{ISL}} \) and \( \epsilon_{\text{PSL}} \) in (\ref{Eqn::P1-0b}) and (\ref{Eqn::P1-0c}) represent the desired worst-case sensing security levels. Since the ISL and PSL constraints cannot be expressed in closed form with respect to \( \{w_n\} \), we first determine the parameters \( (p,q,\kappa) \) to shape the secure ACF such that the sensing security constraints are satisfied.

Under the given constellation set for data symbol modulation, the constraints (\ref{Eqn::P1-0b}) and (\ref{Eqn::P1-0c}) on sensing security are reformulated as constraints on \( \{w_n\} \) for the secure ACF. 
Firstly, it is straightforward to determine the sparse subcarrier spacing \( \kappa \) in relation to \( \epsilon_{\text{ISL}} \) and \( \epsilon_{\text{PSL}} \). By setting \( \Delta_{\text{PSL,E}} = \epsilon_{\text{PSL}} \) and substituting (\ref{eq30}) into (\ref{Eqn::P1-0b}), we obtain
\begin{align}
    \kappa \geq \frac{\epsilon_{\text{ISL}} - \mu_4 + 1}{\epsilon_{\text{PSL}} \mu_4} +1. \label{eq36}
\end{align}
Suppose that the number of subcarriers \( N \) is a power of 2. Then, the smallest \( \kappa \) satisfying (\ref{eq36}) can be determined as
\begin{align}
    \kappa = 2^{\left[\log_2 \left(\frac{\epsilon_{\text{ISL}} - \mu_4 + 1}{\epsilon_{\text{PSL}} \mu_4} +1\right)\right]+1}. \label{eq37}
\end{align}
Furthermore, we replace (\ref{Eqn::P1-0c}) using (\ref{eq22}) and (\ref{eq27}) with 
\begin{align}
    \sum_{n\in \mathcal{P}_{n_0,\kappa}^c} |w_n|^2 \leq \frac{N(\kappa-1)}{\kappa}(1 - \sqrt{\epsilon_{\text{PSL}}}). \label{eq38}
\end{align}
Substituting (\ref{eq37}) and (\ref{eq38}) into (\ref{Eqn::P1-0b}) and (\ref{Eqn::P1-0c}), we reformulate problem \((\mathbf{P.1})\) as
\begin{align}
    (\mathbf{P.2}) \; & \underset{\{w_n\}}{\text{maximize}}
    \qquad   -(1-\rho) \frac{\mathcal{L}_{A}}{\mathcal{L}_{A,\rho=0}} + \rho \frac{R_c}{R_{c,\rho=1}} \label{Eqn::P2-0a} \\
    & \text{subject to} 
    \qquad \qquad    \text{(\ref{eq37}), (\ref{eq38}), and (\ref{Eqn::P1-0d})} \nonumber.
\end{align}
Referring to (\ref{eq5}) and Theorem \ref{theo2}, the formulated problem is a convex quadratic program, which can be efficiently solved using numerical convex optimization tools.

\section{Numerical Results}
In this section, we present numerical simulation results to validate the proposed sensing-secure ISAC framework. Unless stated otherwise, the ISAC TX transmits an OFDM signal with \( B = 50 \) MHz, \textcolor{black}{\( N = 256 \), and $N_{\text{cp}} = 64$, where $R_{\text{max}}=768$ m and $R_{\text{max,cp}}=192$ m. The random communication data is modulated from a 16QAM constellation}, where \( \mu_4 = 1.32 \) and \( \nu_{-2} = 1.89 \). We integrate over 32 OFDM symbols to evaluate the sensing performance at both Alice and Eve. For clarity in graphical illustration, the target range observed by Alice is assumed to be identical to that observed by Eve. Each simulation result is obtained from 1000 Monte Carlo runs.

\subsection{Design of Secure ACF for Artificial Target Generation} \label{sim1}
\begin{figure}[t!]
    \centering
    {\includegraphics[scale=0.7]{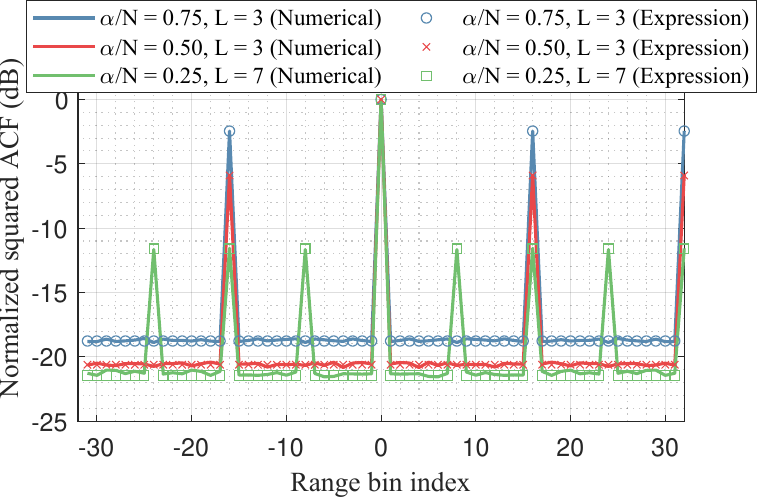}}
    \caption{Various secure ACFs with a 16QAM constellation and \( N = 64 \).}
    \label{f4}
\end{figure}

\begin{figure}[t!]
    \centering
    {\includegraphics[scale=0.7]{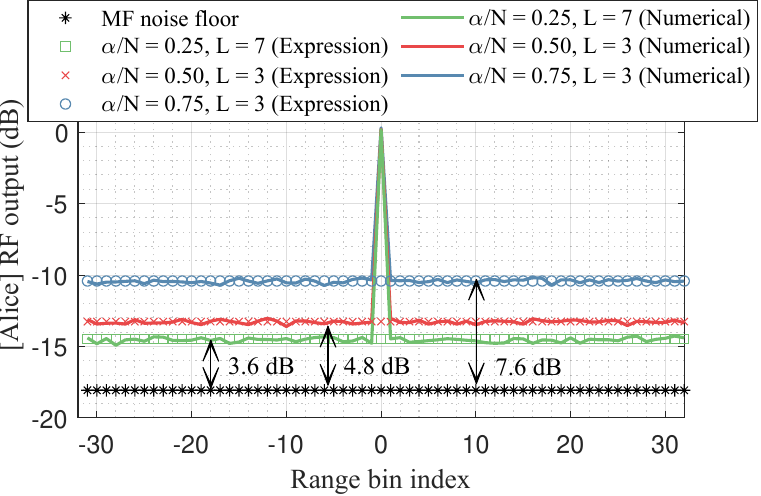}}
    \caption{Alice’s range profiles with RF under the ISAC signals with the secure ACFs in Fig. \ref{f4}. A target is assumed to be located at zero delay with an SNR of 0 dB. The SNR of the MF output is 18 dB.}
    \label{f5}
\end{figure}

Firstly, we present various secure ACFs utilized for sensing-secure ISAC transmission and analyze their corresponding impacts on Alice's RF receiver. Fig. \ref{f4} illustrates the secure ACFs with different numbers and magnitudes of artificial peaks, designed based on Theorem \ref{theo1}. For graphical clarity, we set \( N = 64 \). Recalling that $\alpha$ is the magnitude of the artificial peaks and $L$ denotes the number of artificial peaks, the secure ACF with \( \alpha/N = 0.75 \) and \( L = 3 \) is obtained using \( p = 3.25 \), \( q = 0.25 \), and \( \kappa = 4 \), resulting in \( \Delta_{\text{PSL,E}} = -2.5 \) dB and \( \Delta_{\text{ISL,E}} = 4 \) dB. In contrast, for \( \alpha/N = 0.5 \) and \( L = 3 \), the lower PSL and ISL values of \( \Delta_{\text{PSL,E}} = -6 \) dB and \( \Delta_{\text{ISL,E}} = 1.17 \) dB indicate weaker sensing security compared to the case with \( \alpha/N = 0.75 \). On the other hand, the secure ACF with \( \alpha/N = 0.25 \) and \( L = 7 \) is designed using \( p = 2.75 \), \( q = 0.75 \), and \( \kappa = 8 \), yielding PSL and ISL values of \( \Delta_{\text{PSL,E}} = -12 \) dB and \( \Delta_{\text{ISL,E}} = -0.5 \) dB, representing the lowest security level among the three ACFs.

Notably, the squared secure ACFs in Fig. \ref{f4} exhibit sidelobes across all range bins. This is due to the random signaling with the 16QAM constellation, which introduces additional sidelobes compared to the unit-amplitude constellation, as stated in Proposition \ref{prop1}. The theoretical expression of the secure ACF in (\ref{eq23}) closely matches the numerical simulation results, confirming that the secure ACF exhibits deterministic artificial peaks from subcarrier power allocation while maintaining a sidelobe floor due to the constellation.

Furthermore, we validate the SNR loss at Alice under various secure ACFs. The use of RF to cancel artificial sidelobes in the secure ACF inherently causes SNR loss compared to MF, and this loss is further influenced by the ISAC signaling design, as described in Theorem \ref{theo2}. Fig. \ref{f5} illustrates the range profiles of Alice's RF output under ISAC signaling with secure ACFs, assuming a single target is located at zero delay. Here, we observe that the resulting range profiles with RF exhibit no artificial peaks. Instead, each signaling scheme with a different secure ACF results in a different level of SNR loss compared to the MF noise floor, 3.6 dB, 4.8 dB, and 7.6 dB, respectively. Interestingly, the SNR loss increases as the security level of the ACF improves, revealing a trade-off between sensing security and legitimate sensing performance. 

From Corollary \ref{corr3}, the theoretical SNR loss is determined by both the subcarrier power allocation and the inverse second moment of the constellation. The numerical simulations confirm this theoretical proof, showing a close match between the theoretical and simulated results. In conclusion, these findings provide valuable insight into the signaling design for sensing-secure ISAC, highlighting the inherent trade-off between introducing strong artificial targets at Eve and incurring SNR loss at Alice.

\subsection{Three-way Trade-offs in Sensing-Secure ISAC Signaling}
\begin{figure}[t!]
    \centering
    \subfloat[]{\includegraphics[scale=0.7]{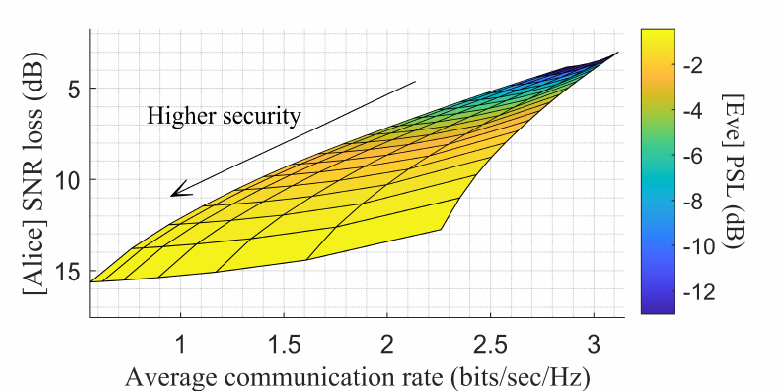}} \\
    \centering
    \subfloat[]{\includegraphics[scale=0.7]{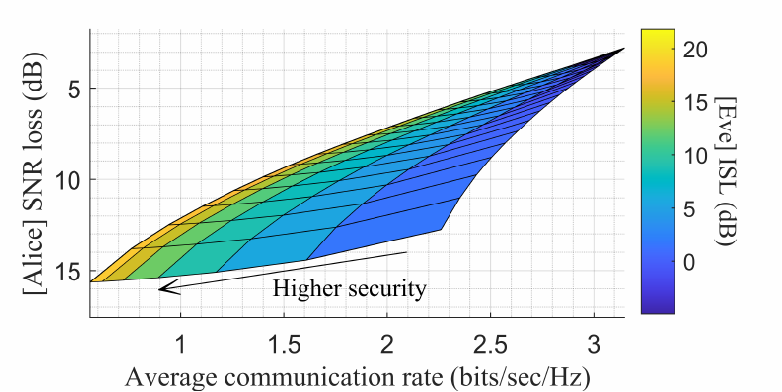}} 
    \caption{\textcolor{black}{Three-way trade-offs between communication, sensing, and sensing security: (a) ISAC vs. PSL of the secure ACF, and (b) ISAC vs. ISL of the secure ACF.}}
    \label{f6}
\end{figure}

In this section, we investigate the three-way trade-offs between communication, sensing, and sensing security in the proposed sensing-secure ISAC framework. As discussed in Section \ref{section4}, the proposed AF shaping approach introduces a new trade-off between ISAC performance and sensing security, with feasible performance points determined by (\ref{eq35}). Without considering S\&C performance optimization, we explore this trade-off by sweeping the secure ACF design parameters \( (p,q,\kappa) \), as shown in Fig. \ref{f6}, where the average communication SNR per subcarrier \( |h_i|^2/\sigma_c^2 \) is set to 10 dB. The minimum SNR loss and maximum communication rate are achieved simultaneously when the PSL and ISL of the ACF are minimized. Conversely, the highest SNR loss and lowest communication rate occur when sensing security is maximized, confirming the inherent trade-off between ISAC performance and sensing security. Here, it is worth noting that the minimum SNR loss at Alice is bounded due to the random signaling nature of 16QAM, as discussed in Section \ref{sim1}.

On the other hand, the sensitivities of S\&C performance variations with respect to PSL and ISL differ between sensing and communication. As shown in Fig. \ref{f6}(a), the SNR loss is highly influenced by the PSL of the secure ACF, while its variation remains smaller than that of communication performance for a fixed PSL value. This insight suggests that legitimate sensing performance is primarily affected by the magnitude of artificial peaks, whereas the communication rate is more sensitive to the number of artificial peaks, as illustrated in Fig. \ref{f6}(b). Nevertheless, both sensing security metrics negatively impact legitimate sensing and communication performance, reinforcing the inherent trade-off between ISAC and sensing security.

\begin{figure}[t!]
    \centering
    {\includegraphics[scale=0.74]{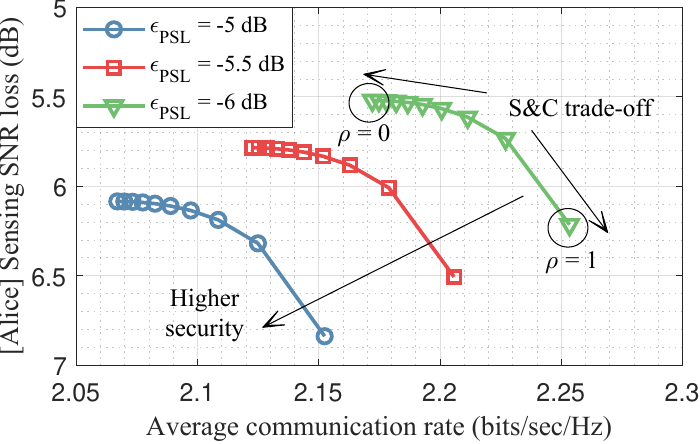}}
    \caption{S\&C trade-offs under given levels of sensing security based on the sensing-secure ISAC signaling design. \( \epsilon_{\text{ISL}} \) is set to 7 dB, resulting in \( \kappa = 16 \).}
    \label{f7}
\end{figure}
Next, we further optimize the sensing-secure ISAC signaling under given PSL and ISL constraints, providing a flexible trade-off between S\&C performance. The key idea behind this optimization is that the optimal subcarrier power allocation for communication can be determined based on CSI, while an increased variation in subcarrier power allocation leads to a higher SNR loss. Fig. \ref{f7} illustrates the S\&C trade-offs under given levels of sensing security, where each secure-sensing ISAC signaling scheme is designed using the proposed framework in Section \ref{signaling}. The results demonstrate that the proposed design enables a flexible trade-off between SNR loss at Alice and the achievable communication rate, controlled by the weighting factor \( \rho \). Once again, these findings validate the inherent trade-off between ISAC performance and sensing security, showing that higher security levels degrade the ISAC performance region, thereby reinforcing the three-way trade-off between communication, sensing, and sensing security.

\begin{figure}[t!]
    \centering
    \subfloat[]{\includegraphics[scale=0.65]{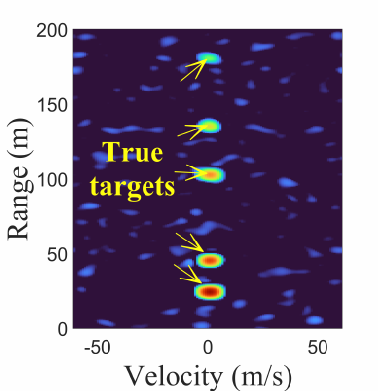}}
    \centering
    \subfloat[]{\includegraphics[scale=0.65]{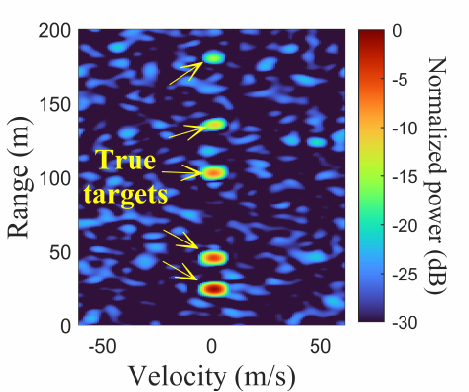}}\\
    \centering
    \subfloat[]{\includegraphics[scale=0.65]{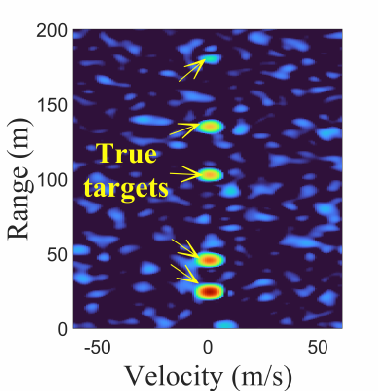}}
    \centering
    \subfloat[]{\includegraphics[scale=0.65]{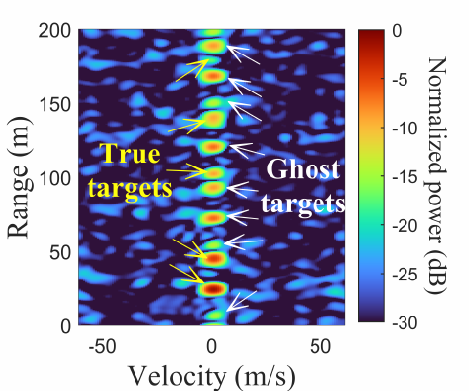}}\\
    \caption{\textcolor{black}{Range-Doppler (RD) maps with five static targets located at different ranges. (a), (b): Alice’s RD maps without and with sensing security, respectively. (c), (d): Eve’s RD maps without and with sensing security, respectively. The sensing-secure ISAC signaling is designed with \( \epsilon_{\text{PSL}} = -5 \) dB, \( \epsilon_{\text{ISL}} = 7 \) dB, and \( \rho = 0 \).}}
    \label{f8}
\end{figure}

\subsection{Security Performance Analysis: Detection and Estimation}
Now, we comprehensively analyze the sensing security performance in terms of target detection and range estimation. Here, we assume that the SINR of the reference signal leaked to Eve is 0 dB. First, we present range-Doppler (RD) maps for both Alice and Eve in the scenario where five static targets \textcolor{black}{$U = U_t = 5$} are located at 24 m, 45 m, 100 m, 135 m, and 180 m, respectively. The targets have SNRs that gradually decrease from 0 dB to -15 dB in steps of -3 dB. Without the secure-sensing ISAC signaling design, both Alice and Eve can clearly detect all targets at zero velocity, as shown in Fig. \ref{f8}(a) and Fig. \ref{f8}(c), respectively. This indicates that Eve can exploit the ISAC signals of opportunity to maliciously detect and estimate targets. On the other hand, the proposed secure ISAC signaling deceives Eve, preventing it from accurately detecting and estimating true targets by introducing artificial peaks in Eve's RD map, as shown in Fig. \ref{f8}(d). Meanwhile, Alice, as shown in Fig. \ref{f8}(b), can still successfully locate the true targets due to the use of RF, albeit at the expense of an SNR loss compared to Fig. \ref{f8}(a).

\begin{figure}[t!]
    \centering
    \includegraphics[scale=0.7]{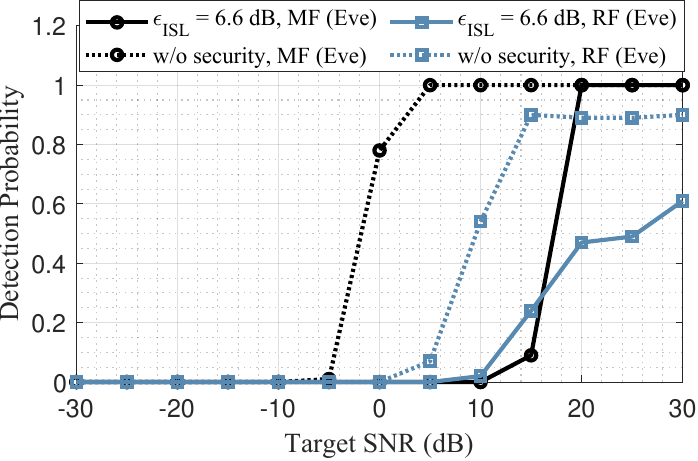}
    \caption{\textcolor{black}{Target detection performance of Eve with reciprocal filtering when clutter with an SNR of 10 dB is located at 30 m, and the target is located at a distance of 100 m. The sensing-secure ISAC signaling is designed with \( \epsilon_{\text{PSL}} = -5 \) dB and \( \rho = 0 \). The SINR of the reference signal leaked to Eve is 0 dB.}}
    \label{f1_R}
\end{figure}

\begin{figure}[t!]
    \centering
    {\includegraphics[scale=0.7]{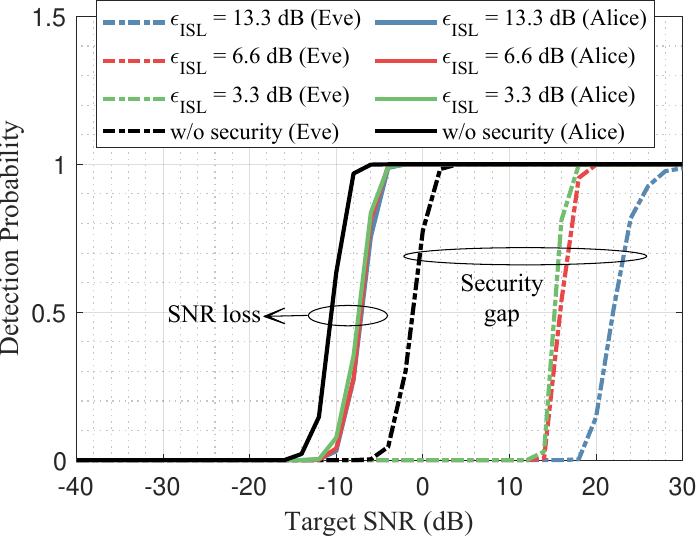}}
    \caption{Target detection performance of Alice and Eve when clutter with an SNR of 10 dB is located at 30 m, and the target is located at a distance of 100 m. The sensing-secure ISAC signaling is designed with \( \epsilon_{\text{PSL}} = -5 \) dB and \( \rho = 0 \).}
    \label{f9}
\end{figure}

We further quantitatively compare the target detection performance between Alice and Eve in the presence of clutter, \textcolor{black}{$U = 2, U_t = 1, U_c = 1$}. Supposing a clutter with an SNR of 10 dB is located at 30 m, we evaluate the target detection performance for a target positioned at a distance of 100 m. A conventional cell-averaging constant false alarm rate (CA-CFAR) detector is employed with a false alarm rate of \(10^{-5}\). \textcolor{black}{First, we evaluate the target detection performance at Eve to support the assumption that Eve employs the MF receiver rather than the RF. As illustrated in Fig.~\ref{f1_R}, the detection performance of Eve with the RF receiver is significantly degraded compared to that with the MF receiver, primarily due to the use of an imperfect reference signal. Furthermore, when the proposed secure-sensing ISAC signaling is transmitted, Eve’s detection performance under the RF receiver deteriorates even further, as the associated SNR loss in the RF receiver becomes more pronounced.}

Fig. \ref{f9} presents the target detection performance of Alice and Eve under various levels of sensing security. The performance of Eve deteriorates significantly as the ISL of the secure ACF increases, as it raises the average power level near the target location, making target detection more challenging by requiring 10 to 100 times higher target power. For the signaling design, the PSL is fixed at -5 dB, meaning that a higher ISL directly corresponds to more artificial peaks in the secure ACF. Compared to Eve, Alice exhibits superior target detection performance. However, the SNR loss results in minor degradation about 2 dB of the required target SNR compared to the case without sensing-secure signaling. Consequently, AF shaping with artificial targets effectively prevents unauthorized Eve from detecting the target while ensuring that legitimate sensing performance remains largely unaffected.

\begin{figure}[t!]
    \centering
    {\includegraphics[scale=0.72]{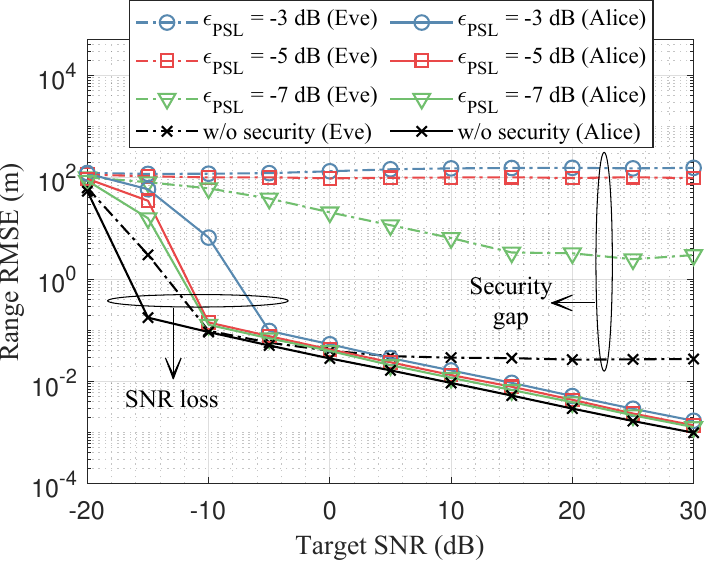}}
    \caption{Range estimation performance of Alice and Eve when two targets have a random SNR difference of 4-6 dB. The sensing-secure ISAC signaling is designed with \( \epsilon_{\text{ISL}} = 7 \) dB, resulting in \( \kappa = 16 \) and \( \rho = 0 \).}
    \label{f10}
\end{figure}

\begin{figure}[t!]
    \centering
    {\includegraphics[scale=0.75]{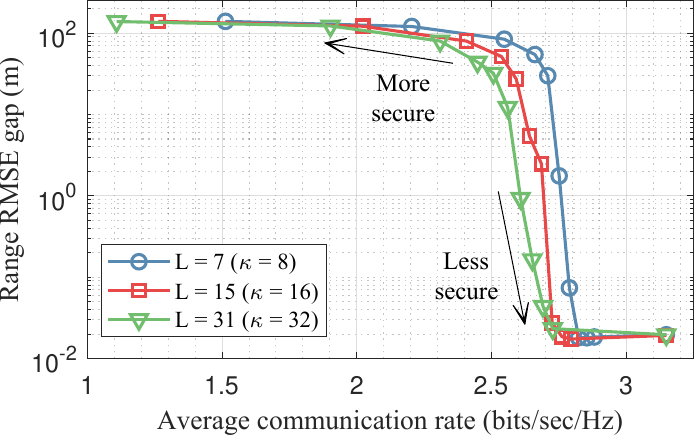}}
    \caption{Range estimation RMSE gap (Eve's RMSE $-$ Alice's RMSE) vs. communication rate when two targets have a random SNR difference of 4-6 dB. The sensing-secure ISAC signaling is designed with varying \( \epsilon_{\text{PSL}}\) and \( \kappa = 8,16,\) and 32, respectively.}
    \label{f11}
\end{figure}
Finally, we explore the multi-target range estimation performance of Alice and Eve. Fig. \ref{f10} illustrates the range estimation performance for two targets, \textcolor{black}{$U = U_t = 2$}, with a random SNR gap of 4–6 dB. The target ranges are estimated using the root MUSIC estimator with a given source number. Compared to the case without sensing security, Eve’s range estimation performance is severely degraded when the sensing-secure ISAC signal is applied, leading to a noticeable security gap more than 100 m estimation errors. This degradation occurs because larger artificial peaks cause Eve to falsely estimate target ranges, resulting in large-scale range estimation errors. Moreover, we define the range RMSE gap between Alice and Eve as a sensing secrecy metric, which characterizes the ultimate trade-off between sensing security performance and communication rate, as illustrated in Fig. \ref{f11}. The results confirm that secure sensing functionality can be flexibly achieved at the expense of communication performance. It is worth noting that introducing more artificial peaks further degrades communication performance to maintain a desired RMSE gap. Nevertheless, the increase of the number of artificial peaks enhances sensing security in terms of target detection performance, as demonstrated in Fig. \ref{f9}. \textcolor{black}{It should be noted that the range RMSE gap narrows when the PSL becomes extremely high, i.e., as $\alpha/N \rightarrow 1$ or equivalently $q \rightarrow 0$ in \eqref{eqCorr3}, because the SNR loss at the RF output, denoted by $\mathcal{L}_A$, increases sharply, thereby degrading legitimate sensing performance. Therefore, it is recommended to avoid selecting a large value of \( \alpha/N \) for secure ISAC signaling to preserve legitimate sensing functionality.} 

\textcolor{black}{Building on the sensing security performance analysis of detection and estimation, we observe that the bottom-line performance, such as target detection probability and range estimation RMSE gap between Alice and Eve, is significantly improved by the proposed secure ISAC signaling design, which leverages intermediate metrics: the PSL and ISL of the AF.} In conclusion, the target detection and estimation performances of Alice and Eve align with our theoretical and numerical results, confirming the effectiveness of the proposed secure-sensing ISAC signaling with artificial targets.

\section{Conclusion}
In this paper, we proposed a sensing-secure ISAC framework that enhances sensing security against Eve, establishing a fundamental basis for sensing-secure ISAC signaling. By leveraging AF shaping, we introduced artificial targets into Eve’s range profile while ensuring that the legitimate sensing receiver, Alice, retains target detection capabilities using reciprocal filtering. A structured subcarrier power allocation scheme was designed to shape the secure ACF, enabling periodic peak insertion to degrade Eve’s range estimation performance. Furthermore, we formulated and optimized the ISAC signaling design to balance ISAC performance and sensing security under given constraints. Simulation results demonstrated the effectiveness of the proposed secure ISAC framework in degrading Eve’s target detection and estimation performance while maintaining reliable sensing at Alice. The findings highlight the three-way trade-offs in sensing-secure ISAC and provide valuable insights into secure ISAC system design. \textcolor{black}{The proposed fundamental framework can be broadly extended to various research areas in ISAC, including secure-sensing ISAC in high-mobility scenario, estimation- and information-theory based secure ISAC signaling design, multi-input multi-output ISAC beamforming, constellation optimization, pulse-shaping with spectral compliance, multi-path exploitation, offering additional degrees of freedom to further enhance sensing security through AF shaping.}

\appendices
\section{Proof of Theorem \ref{theo1}}
\label{proof_theo1}
The subcarrier power allocation for the secure ACF can be directly obtained by taking DFT over (\ref{eq19}), which is given by
\begin{align}
    |w_n|^2 & = \frac{1}{N}\sum_{k=1}^{N} \tilde{\Lambda}[k]e^{-j\frac{2\pi}{N}k(n-1)}  \nonumber \\ 
    & = 1 + \frac{\alpha}{N} \sum_{l=1}^{L} e^{-j\frac{2\pi}{N}l\lambda (n-1)} , \forall{n = 1, 2, \dots,N}. \label{eqA1} 
\end{align}
With $L+1 = N/\lambda$ in hand, it is further simplified as
\begin{align}
    |w_n|^2 & = \left(1-\frac{\alpha}{N} \right) + \frac{\alpha}{N} \sum_{l=1}^{L+1} e^{-j\frac{2\pi l}{(L+1)} (n-1)} \nonumber \\
    & = \left(1-\frac{\alpha}{N} \right) + \frac{\alpha (L+1)}{N} \sum_{d=0}^{\lambda-1} \delta[n-1-\kappa d], \quad \forall{n}, \label{eqA2}
\end{align}
where $\kappa = L+1$. Thus, the power allocation of subcarriers for the secure ACF is equivalently written as a form of (\ref{eq20}) with the dominant power allocation set $\mathcal{P}= \left\{ 1,1+\kappa,\dots \right\}.$

\section{Proof of Corollary \ref{corr1}}
\label{proof_corr1}
The circular shift of (\ref{eqA2}) by $n_0$ samples is expressed as
\begin{align}
    |w_n|^2 & = \left(1-\frac{\alpha}{N} \right) + \frac{\alpha (L+1)}{N} \sum_{d=0}^{\lambda-1} \delta[n-n_0-1-\kappa d]. \label{eqA3}
\end{align}
Then, its ACF is given by a linear phase-shifted form of the secure ACF in (\ref{eq19}), which is written as
\begin{align}
    {\Lambda}_{n_0}[k] = \left(N{\delta[k]} + \alpha \sum_{l=1}^{L} \delta[k-l\lambda]\right) e^{-j\frac{2\pi}{N}k n_0}, \quad \forall{k}.
\end{align}
Thus, the squared ACF $\left|{\Lambda}_{n_0}[k]\right|^2$ remains same to the squared secure ACF $\left|\tilde{\Lambda}[k]\right|^2$ of (\ref{eq19}), completing the proof.

\section{Proof of Corollary \ref{corr2}}
\label{proof_corr2}
From (\ref{eq18}), the expected squared ACF is given as
\begin{align}
    & \mathbb{E}\left[|{\Lambda}[k]|^2\right] = \mathbb{E}\left[\sum_{n=1}^N \sum_{m=1}^N |w_n|^2|w_m|^2e^{j\frac{2\pi}{N}k(n-m)}\right] \nonumber \\
    & = \sum_{n = 1}^N \mathbb{E}\left[|w_n|^4\right]  +  \mathbb{E}\left[\sum_{n=1}^N \sum_{m \neq n}^N |w_n|^2|w_m|^2 e^{j\frac{2\pi}{N}k(n-m)}\right] \nonumber \\
    & = \sum_{n = 1}^N \left[\mathbb{E}\left[|w_n|^4\right] -  \left(\mathbb{E}\left[|w_n|^2\right]\right)^2\right] \nonumber \\ 
    & \qquad \quad + \sum_{n=1}^N \sum_{m =1}^N \mathbb{E}\left[|w_n|^2\right]\mathbb{E}\left[|w_m|^2\right]e^{j\frac{2\pi}{N}k(n-m)}.
    \label{eqCorr2}
\end{align}
Based on (\ref{eq22}) and Theorem \ref{theo1}, we have 
\begin{align}
    \mathbb{E}\left[|{\Lambda}[k]|^2\right] = \sum_{n = 1}^N \text{Var}\left[|w_n|^2\right] + \left| \tilde{\Lambda}[k]\right|^2 \approx \left| \tilde{\Lambda}[k]\right|^2,
\end{align}
where $\text{Var}\left[|w_n|^2\right]$ is sufficiently small. Thus, this completes the proof.

\section{Proof of Proposition \ref{prop1}}
\label{proof_prop1}
From (\ref{eq18}), we derive the expected squared ACF as
\begin{align}
    & \mathbb{E}\left[|{\Lambda}[k]|^2\right] \nonumber \\
    & = \mu_{4}\sum_{n = 1}^N |w_n|^4  +  \mathbb{E}\left[\sum_{n=1}^N \sum_{m \neq n}^N |w_n|^2|w_m|^2|s_n|^2|s_m|^2 e^{j\frac{2\pi}{N}k(n-m)}\right] \nonumber \\
    & = (\mu_{4}-1)\sum_{n = 1}^N |w_n|^4 + \sum_{n=1}^N \sum_{m =1}^N |w_n|^2|w_m|^2e^{j\frac{2\pi}{N}k(n-m)}. \label{eqA4}
\end{align}
By substituting (\ref{eq20}) into (\ref{eqA4}), the first term is written as
\begin{align}
    (\mu_{4}-1)\sum_{n = 1}^N |w_n|^4 = (\mu_{4}-1) \left(\frac{N}{\kappa}p^2 + N\left(1 - \frac{1}{\kappa}\right)q^2 \right). \label{eqA5}
\end{align}
With Theorem \ref{theo1} at hand and plugging (\ref{eqA5}) into (\ref{eqA4}) leads to (\ref{eq23}), completing the proof.

\section{Proof of Lemma \ref{lemma1}}
\label{proof_lemma1}
By the definition of the SNR in (\ref{eqSNR}), the output SNR of the MF from (\ref{eq9}) is given as
\begin{align}
    \gamma_{MF} = \frac{\mathbb{E}\left[\beta_{A} \sum_{n=1}^N |w_n|^2|s_n|^2 / \sqrt{N}\right]^2}{\sigma_A^2} =N \beta_{A}^2\sigma_{A}^{-2}.
\end{align}
Similarly, the output SNR of the RF from (\ref{eq11}) is written by
\begin{align}
    \gamma_{RF} = \frac{N\beta_{A}^2 }{\mathbb{E}[|\tilde{z}_{A,RF,s}[n]|^2]},
\end{align}
where the denominator can be further expanded as
\begin{align}
    \mathbb{E}[|\tilde{z}_{A,RF,s}[n]|^2] & = \frac{1}{N} \sum_{n=1}^N \mathbb{E}\left[ \frac{|z_{A,s}[n]|^2}{|w_n|^2 |s_n|^2} \right] \nonumber \\
    & = \frac{\sigma_A^2}{N} \sum_{n=1}^N |w_n|^{-2} \mathbb{E}\left[ |s_n|^{-2} \right].
\end{align}
By the definition of the inverse second moment in (\ref{eq1}), the SNR of RF is provided as (\ref{eq26}).
\ifCLASSOPTIONcaptionsoff
  \newpage
\fi



\bibliographystyle{IEEEtran}
\bibliography{IEEEabrv,reference}
%



%

\begin{IEEEbiography}[{\includegraphics[width=1in,height=1.25in,clip,keepaspectratio]{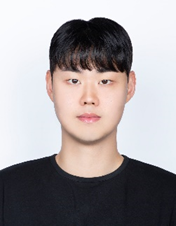}}]{\bf{Kawon Han}\;}  received his B.S., M.S., and Ph.D. degrees in electrical engineering from Korea Advanced Institute of Science and Technology, Daejeon, South Korea, in 2017, 2019, and 2022, respectively. In 2023, he was a postdoctoral researcher at the Interuniversity Microelectronics Centre (IMEC), Leuven, Belgium. From 2024, he joined the Department of Electronics and Electrical Engineering, University College London (UCL), U.K. as a Marie Skłodowska Curie Actions (MSCA) Postdoctoral Fellow. He served as the Track Co-Chair of the IEEE VTC-spring 2025 and a TPC member of the IEEE GlobeCom 2024 Workshop on Integrated Sensing and Communications for Edge Intelligence. His current research interests include integrated sensing and communication systems, distributed and network-level ISAC, ISAC security, microwave/RF system design, biomedical radar signal processing, array signal processing, and millimeter-wave imaging radar.
\end{IEEEbiography}

\begin{IEEEbiography}[{\includegraphics[width=1in,height=1.25in,clip,keepaspectratio]{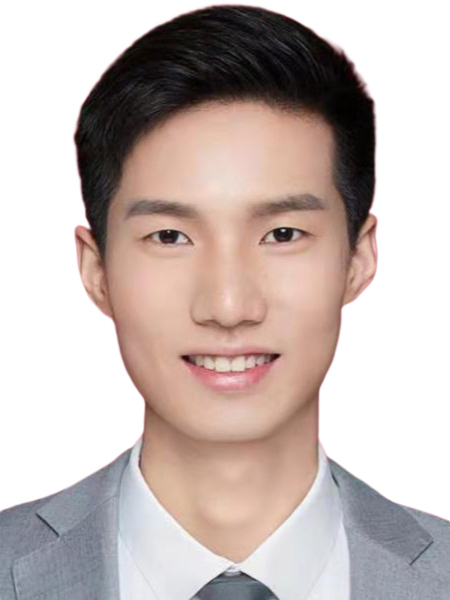}}]{\bf{Kaitao Meng}\;} received the B.E. and the Ph.D. degrees from the School of Electronic Information, Wuhan University, Wuhan, China, in 2016 and 2021, respectively. He is currently working as a Lecturer (Assistant Professor) in the Department of Electrical and Electronic Engineering at the University of Manchester, UK. From 2021 to 2023, he was a Postdoctoral Researcher at the State Key Laboratory of Internet of Things for Smart City, University of Macau, Macau, China. From 2023 to 2025, he worked as a Marie Curie Fellow in the Department of Electronic and Electrical Engineering, University College London, UK. He was the recipient of 2022 EU Marie Curie Postdoctoral Fellowship. He serves as an Editor for IEEE Transactions on Communications, Associate Editor for IEEE Internet of Things Journal, and a Guest Editor for IEEE Transactions on Cognitive Communications and Networking. He has chaired various tracks and workshops at IEEE conferences, including Track Co-Chair of IEEE VTC-Spring 2025 and Workshop Co-Chair of IEEE ICC 2025, IEEE GlobeCom 2024 and 2025, and WiOpt 2025. His current research interests mainly include network-level integrated sensing and communication (ISAC), cooperative sensing and communication, intelligent surfaces, and multi-UAV collaboration.
\end{IEEEbiography}

\begin{IEEEbiography}[{\includegraphics[width=1in,height=1.25in,clip,keepaspectratio]{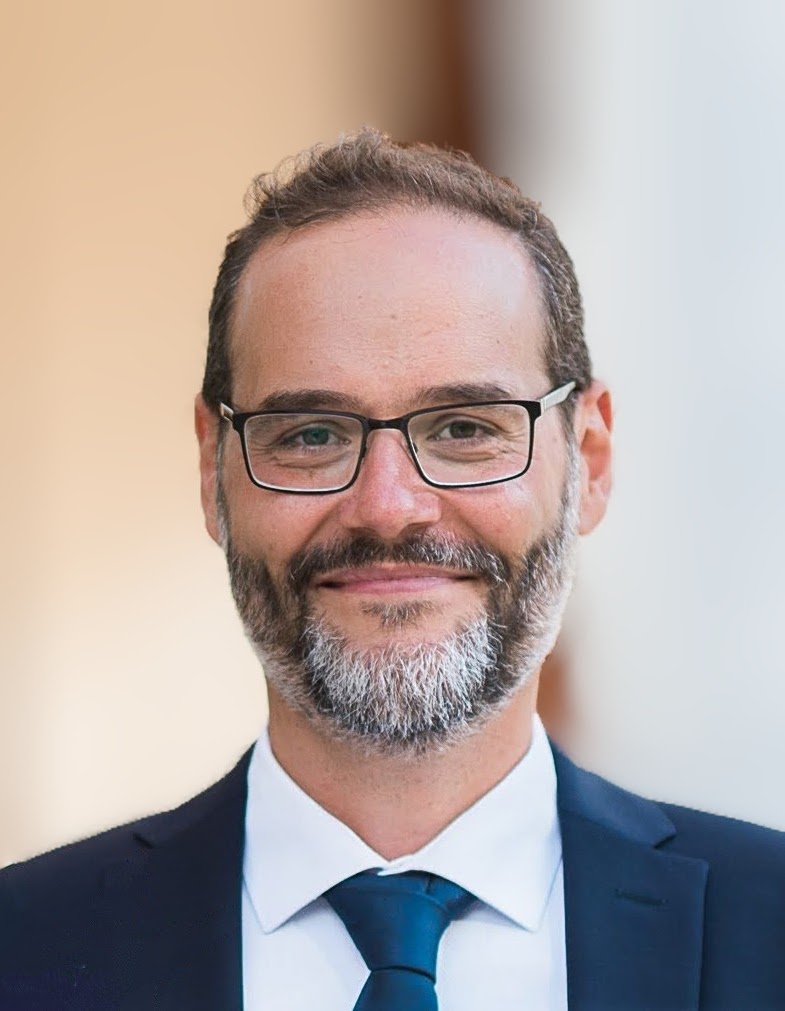}}]{\bf{Christos Masouros}\;}  (Fellow, IEEE) received the Diploma degree in electrical and computer engineering from  University of Patras, Greece, in 2004, and the M.Sc. by research and Ph.D. degrees in electrical and electronic engineering from The University of Manchester, U.K., in 2006 and 2009, respectively. In 2008, he was a Research Intern at Philips Research Labs, U.K. Between 2009 and 2010, he was a Research Associate with The University	of Manchester and a Research Fellow at Queen’s University Belfast between 2010 and 2012. In 2012, he joined University	College London as a Lecturer. He has held a Royal Academy of Engineering	Research Fellowship between 2011 and 2016. Since 2019, he has been a Full Professor of signal processing and wireless communications with the Information and Communication Engineering Research Group, Department of Electrical and Electronic Engineering, and affiliated with the Institute for Communications and Connected Systems, University College London.	His research interests lie in the field of wireless communications and signal processing with particular focus on green communications, large scale antenna systems, integrated sensing and communications (ISAC), interference mitigation techniques for MIMO, and multicarrier communications. He was the co-recipient of the 2021 IEEE SPS Young Author Best Paper Award, and the recipient of the Best Paper Awards in the IEEE GlobeCom	2015 and IEEE WCNC 2019 conferences. He has been recognized as an Exemplary Editor for IEEE Communications Letters and as an Exemplary Reviewer for  IEEE Transactions on Communications. He is a founding member and the Vice-Chair of the IEEE Emerging Technology Initiative on ISAC, the Vice Chair of the IEEE Special Interest Group on ISAC, and the Chair of the IEEE Special Interest Group on Energy Harvesting Communication Networks. He is an Editor of IEEE Transactions on Communications, IEEE Transactions on Wireless Communications, IEEE Open Journal of Signal Processing, and the Editor-at-Large of IEEE Open Journal of the Communications Society. He was an Associate Editor of IEEE Communications Letters and a Guest Editor for a number of special issues on IEEE Journal of Selected Topics in Signal Processing and IEEE Journal on Selected Areas in Communications.
\end{IEEEbiography}





\vfill


\end{document}